\documentclass[ALICE,manyauthors]{cernphprep}

\usepackage[comma,square,numbers,sort&compress]{natbib}
\usepackage{hyperref}
\usepackage{url}
\usepackage{lineno}
\usepackage{epsfig} 
\usepackage{epstopdf} 
\usepackage{comment}
\usepackage{amsmath}
\usepackage{amsthm}
\usepackage{amsfonts}
\usepackage{xspace}
\usepackage{datetime}
\usepackage{mathtools}
\usepackage{mathrsfs}
\usepackage[T1]{fontenc}

\usepackage[english,colorinlistoftodos]{todonotes}
\presetkeys{todonotes}{inline, color=blue!30}{}

\newcommand{\jpsi}{\ensuremath{\text{J}/\psi}\xspace}
\newcommand{\pp}{\ensuremath{\rm pp}\xspace}

\newcommand{\PbPb}{\ensuremath{\text{Pb--Pb}}\xspace}
\newcommand{\XeXe}{\ensuremath{\text{Xe--Xe}}\xspace}
\newcommand{\s}{\ensuremath{\sqrt{s}}\xspace}
\newcommand{\snn}{\ensuremath{\sqrt{s_{\rm NN}}}\xspace}

\newcommand{\pt}{\ensuremath{p_{\text{T}}}\xspace}
\newcommand{\pT}{\ensuremath{p_{\text{T}}}\xspace}
\newcommand{\mee}{\ensuremath{m_{\rm ee}}\xspace}

\newcommand{\der}{\ensuremath{\text{d}}\xspace}

\newcommand{\ccBar}{\ensuremath{\text{c}\overline{\text{c}}}\xspace}
\newcommand{\bbBar}{\ensuremath{\text{b}\overline{\text{b}}}\xspace}
\newcommand{\dEdx}{\ensuremath{\text{d}E/\text{d}x}\xspace}
\newcommand{\RAA}{\ensuremath{R_{\text{AA}}}\xspace}

\newcommand{\TAA}{\ensuremath{T_{\text{AA}}}\xspace}
\newcommand{\Raa}{\ensuremath{R_{\text{AA}}}\xspace}
\newcommand{\raa}{\ensuremath{r_{\text{AA}}}\xspace}

\newcommand{\diele}{\ensuremath{{\rm e}^{+}{\rm e}^{-}}\xspace}

\newcommand{\AccEff}{\ensuremath{A\times\epsilon} \xspace}

\newcommand{\GeVc}{\ensuremath{{\rm GeV}/c}\xspace}

\newcommand{\GeVcsq}{\ensuremath{{\rm GeV}/c^2}\xspace}
\newcommand{\MeVcsq}{\ensuremath{{\rm MeV}/c^2}\xspace}

\newcommand{\MeanPt}{\ensuremath{\langle p_{\rm T} \rangle} \xspace}
\newcommand{\MeanPtSq}{\ensuremath{\langle p_{\rm T}^{2} \rangle} \xspace}

\newcommand{\MeanPtee}{\ensuremath{\langle p_{\rm T} ^{\rm ee} \rangle} \xspace}

\newcommand{\AvgTAA}{\ensuremath{\langle T_{\text{AA}} \rangle}\xspace}
\newcommand{\AvgNpart}{\ensuremath{\langle N_{\rm part} \rangle} \xspace}
\newcommand{\AvgNcoll}{\ensuremath{\langle N_{\rm coll} \rangle} \xspace}
\newcommand{\npart}{\ensuremath{N_{\rm part}} \xspace}
\newcommand{\Npart}{\ensuremath{N_{\rm part}} \xspace}

\begin{document}%

\begin{titlepage}
\PHyear{2019}
\PHnumber{234}      
\PHdate{25 October}  
%


\title{Centrality and transverse momentum dependence of inclusive \jpsi production at midrapidity in Pb--Pb collisions at $\snn=5.02$~TeV}
\ShortTitle{\jpsi nuclear modification factor in Pb--Pb collisions at $\snn=5.02$~TeV}   

\Collaboration{ALICE Collaboration\thanks{See Appendix~\ref{app:collab} for the list of collaboration members}}
\ShortAuthor{ALICE Collaboration} 

\begin{abstract}
  The inclusive $\jpsi$ meson production in $\PbPb$ collisions at a center-of-mass energy per nucleon--nucleon collision of $\snn = 5.02$~TeV at midrapidity ($|y| < 0.9$) is reported by the ALICE Collaboration. The measurements are performed in the dielectron decay channel, as a function of event centrality and $\jpsi$ transverse momentum $\pT$, down to $\pT=0$. The $\jpsi$ mean transverse momentum $\MeanPt$ and $\raa$ ratio, defined as $\MeanPtSq_{\text{PbPb}} / \MeanPtSq_{\mathrm{pp}}$, are evaluated. Both observables show a centrality dependence decreasing towards central (head-on) collisions. The $\jpsi$ nuclear modification factor $\RAA$ exhibits a strong $\pT$ dependence with a large suppression at high $\pt$ and an increase to unity for decreasing $\pt$. When integrating over the measured momentum range $\pT<10$~\GeVc, the $\jpsi$ $\RAA$ shows a weak centrality dependence. Each measurement is compared with results at lower center-of-mass energies and with ALICE measurements at forward rapidity, as well as to theory calculations. All reported features of the $\jpsi$ production at low $\pt$ are consistent with a dominant contribution to the $\jpsi$ yield originating from charm quark (re)combination.

\end{abstract}
\end{titlepage}
\setcounter{page}{2}

\section{Introduction}
\label{sec:Introduction}

The Quark-Gluon Plasma (QGP) is a state of strongly-interacting matter characterized by quark and gluon degrees of freedom predicted by Quantum Chromodynamics (QCD) to exist at high temperature and energy density~\cite{PhysRevD.80.014504,Borsanyi:2013bia}. Such conditions are realized during the initial hot and dense stages of ultra-relativistic heavy-ion collisions. The medium produced in these collisions has a short lifetime, which is of the order of $10$~fm/$c$ at the energies reached at the CERN Large Hadron Collider (LHC), see e.g.~\cite{Acharya:2017qtq}.

Due to their large masses, charm and beauty quarks are produced in hard partonic scatterings occurring during the early stage of the collision and therefore experience the full evolution of the medium. Charmonia, i.e.\ the bound states of charm and anti-charm quarks, are of particular interest for the understanding of the QGP, see e.g.~\cite{Andronic:2015wma,Mocsy:2013syh}. In the framework of color-screening models, the suppression of the charmonium state $\jpsi$ is an unambiguous signature of the QGP~\cite{Matsui:1986dk,Karsch:2005nk}. The high density of color charges prevents charm and anti-charm quarks from forming bound states. Therefore, the $\jpsi$ yield is expected to be suppressed compared to probes unaffected by the hot and dense medium or from expectations of the incoherent superposition of nucleon--nucleon collisions at the same energy. This was experimentally observed in the most central heavy-ion collisions at SPS~\cite{Baglin:1994ui,Alessandro:2004ap,PhysRevLett.99.132302} and RHIC~\cite{Adare:2006ns,Adare:2011yf,Adamczyk:2013tvk} energies.

At the significantly higher collision energies of the LHC, the suppression pattern of $\jpsi$ mesons in heavy-ion collisions is fundamentally changed. In central $\PbPb$ collisions at $\snn=~2.76$~TeV, where $\snn$ is the center-of-mass collision energy per nucleon--nucleon pair, the suppression was found to be weaker~\cite{Abelev:2013ila,Abelev:2012rv,Adam:2015isa} in comparison with the earlier measurements at lower energies mentioned above. The effect was measured by the ALICE Collaboration at both mid- and forward rapidity, dominantly for $\jpsi$ mesons at a low transverse momentum ($\pt$). This phenomenon is understood as the result of the charmonium (re)generation due to copiously produced charm quarks, made possible by the deconfined nature of the QGP. In addition to the weaker nuclear suppression of charmonia, recent observations of non-zero elliptic flow of D~\cite{Sirunyan:2017plt,Acharya:2017qps} and $\jpsi$~\cite{Acharya:2017tgv} mesons, suggest that charm quarks may thermalize and flow with the bulk particles during the QGP phase.

There are different phenomenological scenarios available for the description of charmonium production in heavy-ion collisions. In the framework of statistical hadronization, all charmonium states are created at chemical equilibrium at the phase boundary and their abundances are determined by thermal weights~\cite{BraunMunzinger:2000px,Andronic:2007bi}. The transport approach considers a continuous production and dissociation of charmonium states already during the QGP phase governed by a set of rate equations~\cite{Thews:2000rj}. Another approach includes charmonium dissociation by the scattering of comoving partons and hadrons with a (re)generation component at LHC collision energies~\cite{Ferreiro:2012rq}. All current models implementing statistical hadronization, microscopic transport approaches~\cite{Zhao:2007hh,Zhou:2014kka} or comover interactions take into account both the hot medium and the cold nuclear matter (CNM)~\cite{Armesto:2006ph} effects mainly originating from the modification of the gluon distribution function in the nucleus compared to the corresponding function of the free nucleon.

In this paper, we present the ALICE measurement of the inclusive $\jpsi$ production at midrapidity in $\PbPb$ collisions at a center-of-mass energy per nucleon pair of $5.02$~TeV. The $\jpsi$ mesons are reconstructed in the central barrel within the rapidity range $|y| < 0.9$ via the $\text{e}^+ \text{e}^-$ decay channel down to $\pT=0$~\GeVc. The $\jpsi$ $\pT$ spectrum is measured in three centrality intervals. The $\jpsi$ average transverse momentum $\MeanPt$ and $\MeanPtSq$ are evaluated as a function of collision centrality: the latter is shown in comparison with the $\jpsi$ $\MeanPtSq$ measured in $\pp$ collisions, via the ratio $\raa  = \MeanPtSq_{\text{PbPb}} / \MeanPtSq_{\pp}$. The nuclear modification factor $\RAA$, which is defined by the ratio of the production yield in $\PbPb$ collisions and the production cross section in $\pp$ collisions normalized by the nuclear overlap function $\AvgTAA$, as a function of event centrality and $\jpsi$ $\pt$, is obtained using the recent ALICE measurement of the inclusive $\jpsi$ cross section in $\pp$ collisions at $\s=~5.02$~TeV~\cite{Acharya:2019lkw}. The new $\pp$ reference and the larger $\PbPb$ data set allow for a significant reduction of the uncertainties compared to our previous measurements at $\snn=~2.76$~TeV~\cite{Abelev:2013ila,Adam:2015rba}. The results are compared with statistical~\cite{BraunMunzinger:2000px}, microscopic parton transport~\cite{Zhao:2007hh,Zhou:2014kka}, and comover~\cite{Ferreiro:2012rq} model calculations. 

The measurements presented in this publication provide a significant complement to results in $\PbPb$ collisions at the same collision energy by the ALICE Collaboration at forward rapidity~\cite{Adam:2016rdg}, the measurements on $\jpsi$ suppression at high $\pT$ by the ATLAS~\cite{Aaboud:2018quy} and CMS~\cite{Sirunyan:2017isk} Collaborations around midrapidity, as well as to results at forward rapidity in $\XeXe$ collisions at $\snn = 5.44$~TeV~\cite{Acharya:2018jvc}.

\section{Apparatus and data sample}
\label{sec:Apparatus}

A detailed description of the ALICE detector and its performance can be found in Refs.~\cite{aliceTDR,performancePaperRun1}. The ALICE central barrel detector allows for high resolution tracking and particle identification over the full azimuthal angle in the pseudorapidity range $|\eta|<0.9$. The entire setup is placed inside a solenoidal magnet, which creates a uniform axial magnetic field of $B = 0.5$~T along the beam direction. 

The main detectors used for the $\jpsi$ meson reconstruction in the $\diele$ decay channel are the Inner Tracking System (ITS)~\cite{ITS} and the Time Projection Chamber (TPC)~\cite{TPC}. The ITS consists of 6 cylindrical layers of silicon detectors placed at radial distances to the beam line from $3.9$~cm to $43$~cm and provides high-precision tracking close to the interaction point as well as the determination of the primary vertex of the event. The two innermost layers form the Silicon Pixel Detector (SPD), the intermediate two layers are the Silicon Drift Detector (SDD), and the outermost layers are the Silicon Strip Detector (SSD).

Placed around the ITS, the TPC detector is a large cylindrical drift chamber extending radially from $85$~cm to $250$~cm from the nominal interaction point ($x=y=z=0$~cm) and longitudinally between $-250$~cm and $+250$~cm. In addition to being the main tracking detector, the TPC also provides particle identification via the measurement of the specific energy loss ($\dEdx$) of charged particles in the detector gas.

The V0 detectors~\cite{VZero} consist of two scintillator arrays, V0A and V0C, which are located on both sides of the nominal interaction point at $z=329$~cm and $z=-90$~cm and cover the pseudorapidity interval $2.8 \leq \eta \leq 5.1$ and $-3.7 \leq \eta \leq -1.7$. The centrality of the events, expressed in fractions of the total inelastic hadronic cross section, is determined via a Glauber fit to the V0 amplitude as described in~\cite{centrality276, centrality502,centralityNote}. In addition, the V0 detectors are used to provide a minimum-bias trigger (MB), defined as the coincidence of signals in both V0 arrays and the beam crossing.

The results presented in this paper are obtained using the MB trigger data collected during the 2015 LHC $\PbPb$ run at a center-of-mass energy per nucleon pair of $5.02$~TeV. Beam-gas events are rejected using timing selections on the signals from the V0 and Zero Degree Calorimeters~\cite{aliceZDC}. Pileup events are rejected online based on V0, but also in the offline analysis based on the correlation between the V0 multiplicity and the number of tracks reconstructed in the TPC. All events must have a reconstructed primary vertex with a longitudinal position within $\pm 10$~cm around the nominal interaction point. Only the events corresponding to the most central $90$\% of the $\PbPb$ inelastic cross section ($0$--$90$\%) are used in this analysis. For these events the MB trigger is fully efficient and the contamination by electromagnetic interactions is negligible. After all selections, a sample of $70$ million events is available for analysis, corresponding to an integrated luminosity of ${L}_{\text{int}} \approx 10~\mu \text{b}^{-1}$.

\section{Analysis methods}
\label{sec:Analysis}

The $\jpsi$ candidates are reconstructed using the $\diele$ decay channel. The selected electron candidates are tracks reconstructed using both the ITS and TPC detectors. They must have a minimum transverse momentum of $1$~\GeVc and pseudorapidity in the range $|\eta|<0.9$. Primary electrons are selected using a maximum distance-of-closest-approach to the event vertex of at most $1$~cm and $3$~cm in the transverse and longitudinal directions, respectively. Additionally, kink-daughters, i.e. secondary tracks from long-lived weak decays of charged particles, are removed from the analysis. In order to improve the resolution of track reconstruction and to reject secondary electrons from photon conversions in the detector material, the tracks are selected to have at least one hit in either of the two SPD layers. Electrons and positrons from photon conversions are further rejected using a prefilter method~\cite{Acharya:2019lkw} in which track candidates forming pairs with an electron-positron invariant mass lower than $50$~\MeVcsq are removed from any further pairing. In the TPC, the electron candidates are required to have at least $70$ out of $159$ possible space points attached to the track, which ensures good tracking and particle identification resolution. 

\begin{figure}[t]
\begin{center}
  \includegraphics[width = 7 cm]{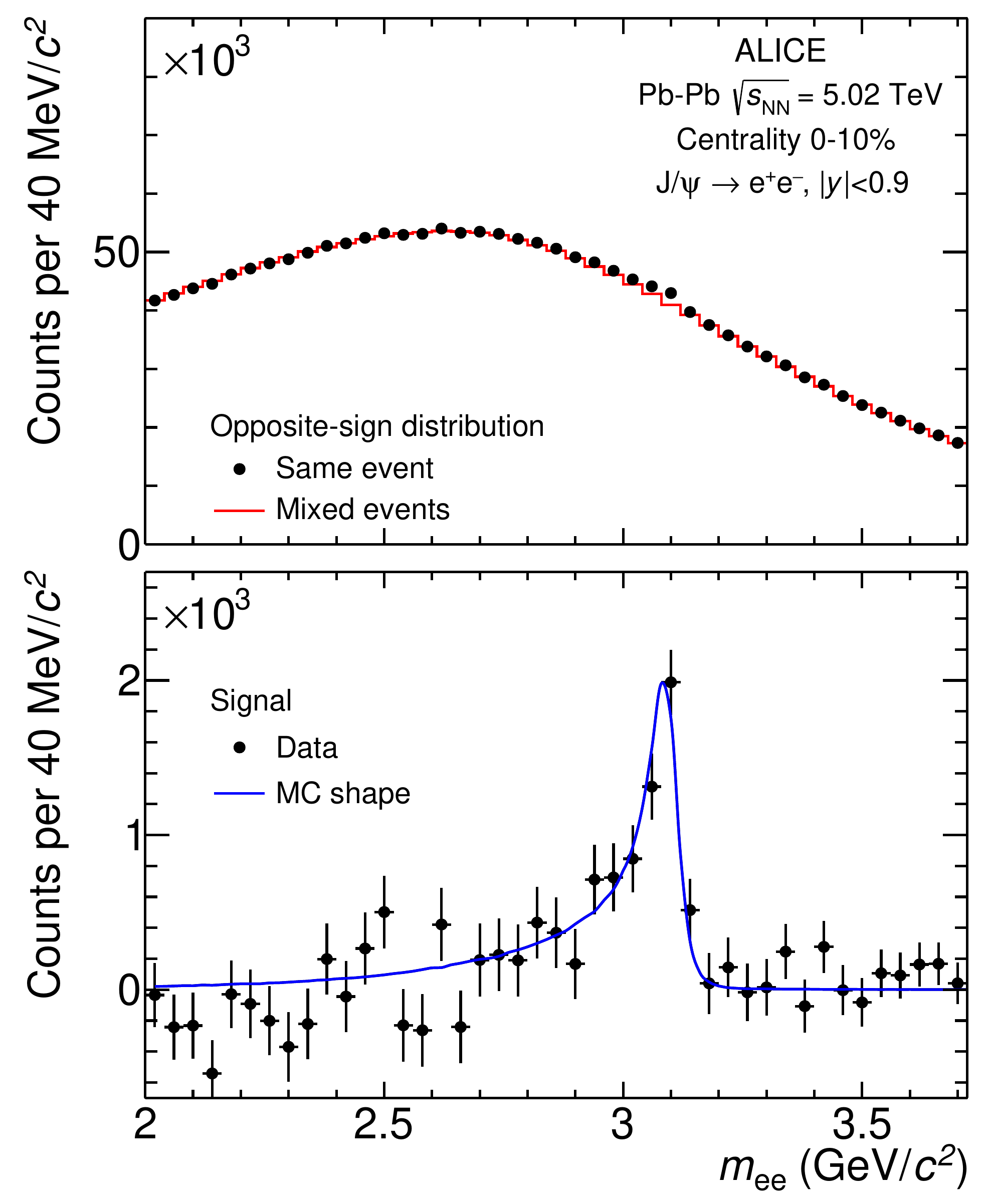}
  \includegraphics[width = 7 cm]{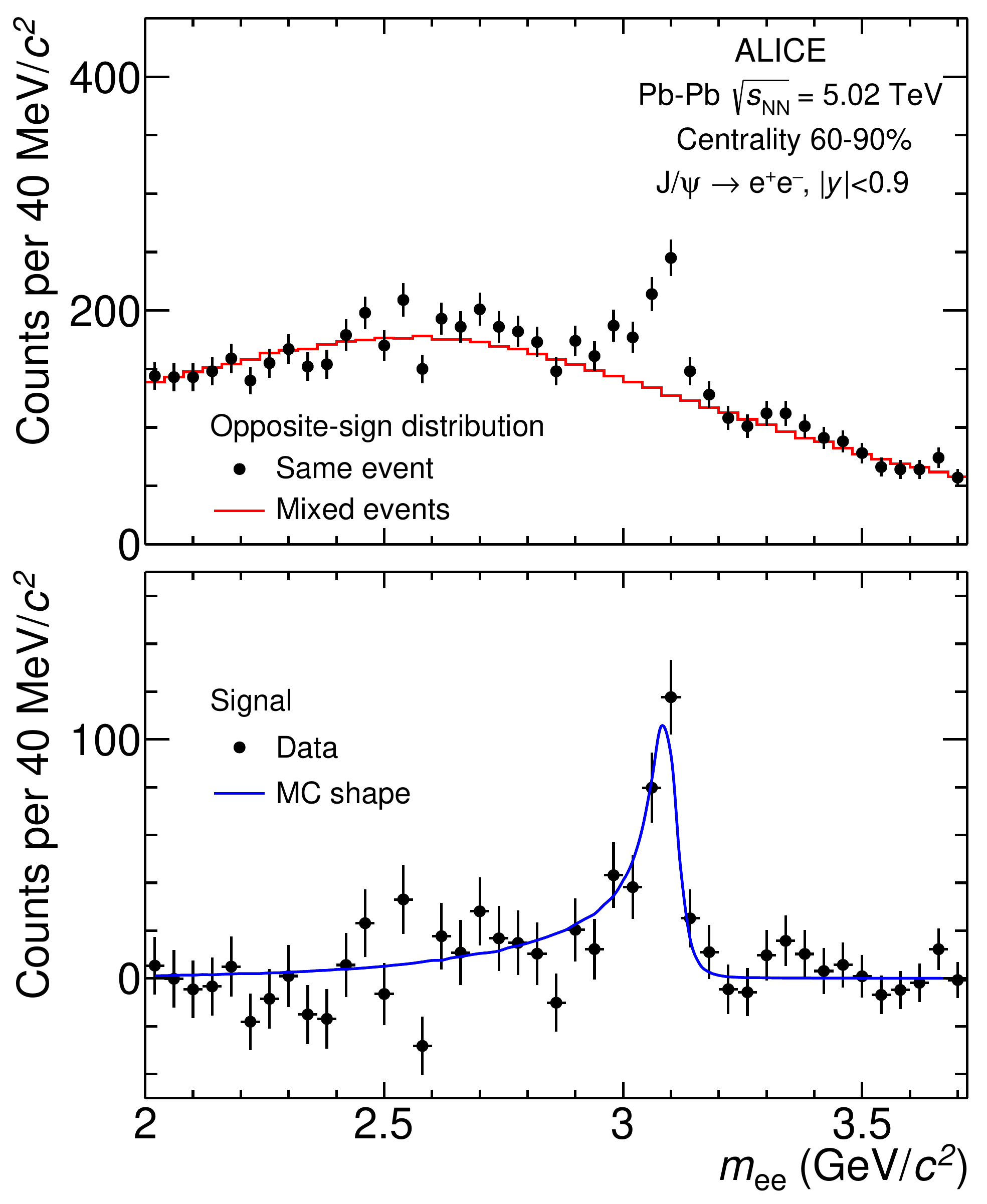}
\caption{(Color online) Top panels: Invariant mass distribution of opposite-sign pairs from the same event and mixed events for the $0$--$10$\% (left) and $60$--$90$\% (right) centrality classes in \PbPb collisions at $\snn = 5.02$~TeV. Bottom panels: Background-subtracted invariant mass distributions in comparison with the expected Monte Carlo signal shape.}
\label{fig:invMass}
\end{center}
\end{figure}

Electrons are identified by requiring that the measured $\dEdx$ in the TPC lies within a $\pm 3\sigma_{\rm{e}}$ band around the expected value for electrons estimated from a parameterization of the Bethe formula, where $\sigma_{\rm{e}}$ is the particle identification (PID) resolution in the TPC for electrons. The hadron contamination is reduced by excluding tracks compatible with the proton or pion hypothesis within $\pm 3.5\sigma_{\rm{p,\pi}}$. 

The number of observed $\jpsi$ is obtained by constructing the invariant mass distribution of all combinations of opposite-sign (OS) electron pairs from the same event. The top panels of Figure~\ref{fig:invMass} show the invariant mass distributions obtained in central ($0$--$10$\%, left) and peripheral ($60$--$90$\%, right) collisions together with the estimated background. The background is obtained using the distribution of OS pairs constructed by pairing electrons and positrons from different events, so-called mixed events (ME), which is scaled to match the same-event OS invariant mass distribution in two mass intervals on either side of the $\jpsi$ signal region: $2.0 <\mee< 2.5~\GeVcsq$ and $3.2 <\mee< 3.7~\GeVcsq$, where the $\jpsi$ contribution is expected to be negligible. The raw $\jpsi$ signal is then obtained by integrating the background-subtracted distribution in the mass window $2.92$--$3.16~\GeVcsq$. The lower panels of Figure~\ref{fig:invMass} show the OS invariant mass distribution after background subtraction. Good agreement with the $\jpsi$ invariant mass distribution from Monte Carlo (MC) simulation, normalized to the integral of the raw signal, is observed. The potentially remaining correlated background from semi-leptonic decays of $\ccBar$ and $\bbBar$ pairs is included in the systematic uncertainty.

The corrected $\jpsi$ $\pt$-differential production yield is obtained in a given centrality class as
\begin{equation}
  \label{eq:yield}
\frac{\der^2 N}{\der y \der \pT} = \frac{N_{\jpsi}}{N_{\text{ev}}\times {\rm BR}_{\jpsi \rightarrow {\rm ee}} \times (\AccEff) \times \Delta y  \times \Delta \pT},
\end{equation}
where $N_{\jpsi}$ is the number of reconstructed $\jpsi$ in the considered centrality class and $\pt$ and $y$ intervals, $N_{\text{ev}}$ is the corresponding number of events, $\AccEff$ is the acceptance and efficiency correction factor and ${\rm BR}_{\jpsi \rightarrow {\rm ee}} = (5.971\pm0.032)$\% is the branching ratio of the $\jpsi$ decaying into the dielectron channel~\cite{Patrignani:2016xqp}.

MC simulations of $\PbPb$ collisions with embedded unpolarized $\jpsi$ mesons are used to obtain the $\AccEff$ factors. The $\PbPb$ collisions are generated using HIJING~\cite{Wang:1991hta}. For the $\jpsi$, the prompt component is generated using a $\pt$ distribution tuned to the existing $\PbPb$ measurements at forward rapidity while the non-prompt component is obtained from $\bbBar$ pairs generated with PYTHIA forced to decay into channels with $\jpsi$ in the final state. The $\jpsi$ decays into the $\diele$ channel are handled using PHOTOS~\cite{photos}. The transport of the simulated particles in the detector material is performed using a GEANT3~\cite{geant3} model of the ALICE apparatus and the same algorithm as for the real data is used to reconstruct the simulated tracks. The acceptance times efficiency correction factors include the kinematic acceptance, the reconstruction and PID efficiencies, and the fraction of signal in the integrated invariant mass window. The acceptance correction factor amounts to 33\% and the fraction of the signal in the mass counting window is approximately 65\%. Reconstruction and PID efficiencies are centrality dependent and together amount to approximately $24$\% in the most central collisions growing monotonically to approximately $32$\% in the most peripheral collisions. The correction factors are also $\pt$ dependent which, for large $\pt$ intervals, induces a dependence on the Monte Carlo $\pt$ distribution of the embedded $\jpsi$. This is taken as a source of systematic uncertainty and is discussed in the following section. The inclusive $\pt$-integrated $\jpsi$ production is measured in $5$ different centrality classes: $0$--$10$\%, $10$--$20$\%, $20$--$40$\% , $40$--$60$\% and $60$--$90$\% while the $\pT$-differential cross sections are obtained in larger centrality classes to ensure sufficient statistical significance: $0$--$20$\%,  $20$--$40$\% and  $40$--$90$\%.

\begin{figure}[t]
\includegraphics[width=0.5\textwidth]{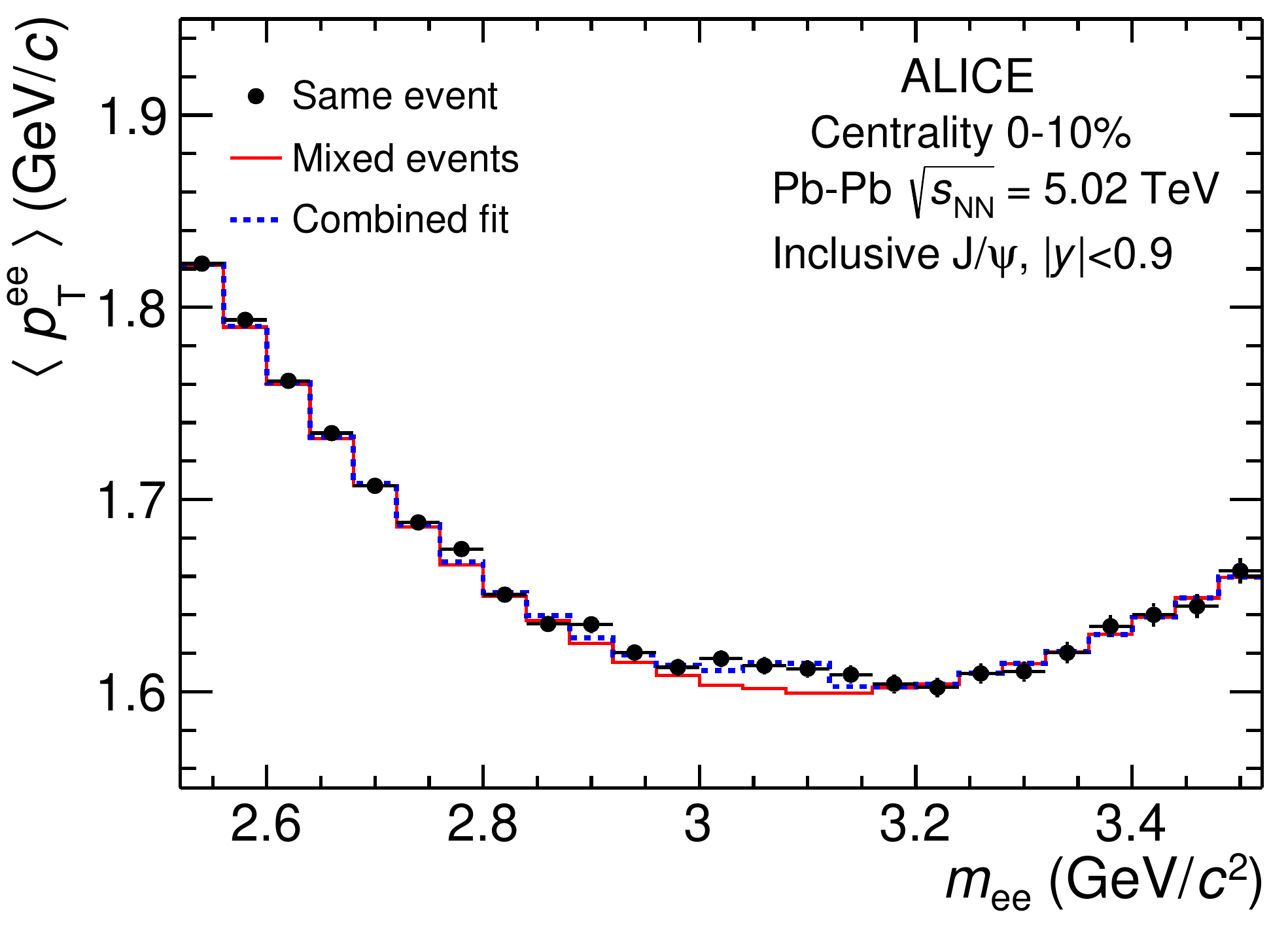}
\includegraphics[width=0.5\textwidth]{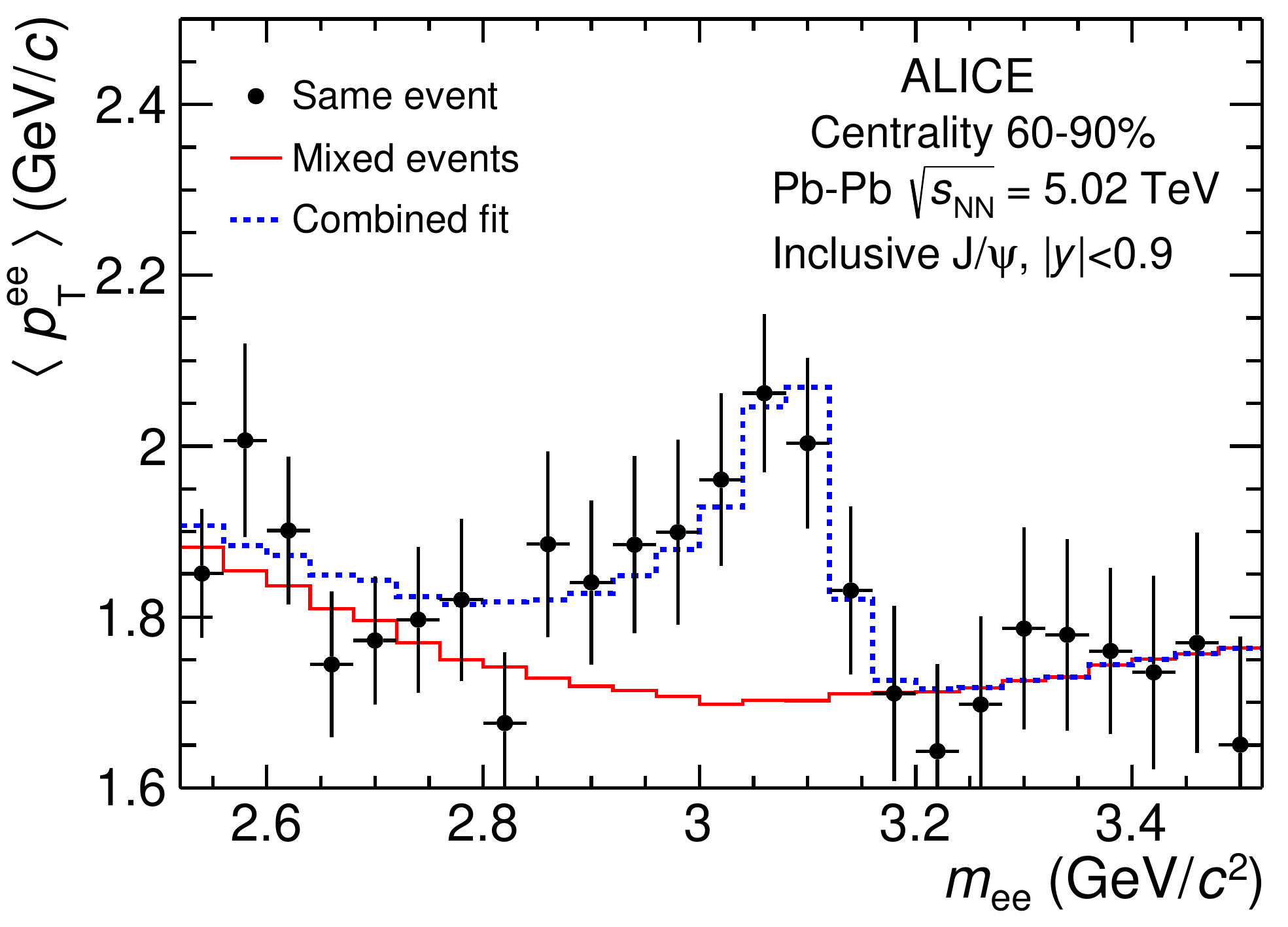}
\caption{(Color online) Extraction of the $\jpsi$ $\MeanPt$ in $\PbPb$ collisions at $\snn = 5.02$~TeV for the $0$--$10$\% (left) and $60$--$90$\% (right) centrality classes in the transverse momentum interval $0.15<\pt<10~$\GeVc. The background, obtained from event-mixing, is shown by the red line.}
\label{meanptextraction}
\end{figure}

The average transverse momentum of $\jpsi$, $\MeanPt$, is extracted using a binned log-likelihood fit of the $\MeanPtee$ distribution of all electron pairs as a function of the invariant mass. Each pair contribution is weighted by the $(\AccEff)^{-1}$ factor corresponding to its centrality and $\pt$. The OS $\MeanPtee$ distribution is fitted with the function:
\begin{equation}
  \MeanPtee (\mee) = \frac{N^{\text{bkg}}(\mee) \times  \langle \pt^{\text{bkg}}(\mee) \rangle + N^{\jpsi}(\mee) \times  \MeanPt }{N^{\rm bkg}(\mee) + N^{\jpsi} (\mee)} ,
  \label{invMassFit}
\end{equation}
where $N^{\rm bkg}(\mee)$ and $N^{\jpsi}(\mee)$ are the mass-dependent distributions of background and signal pairs determined via the signal extraction procedure described above. The background mean transverse momentum, $\langle \pt^{\text{bkg}} \rangle$, depends on the invariant mass and its shape is obtained from the ME technique, while its overall normalization can vary in the fit. Figure~\ref{meanptextraction} illustrates the $\MeanPt$ extraction procedure for the most central and most peripheral centrality intervals using dielectron pairs in the transverse momentum interval $0.15<\pt<10~$~\GeVc. A similar procedure is employed also for the second moment of the transverse momentum distribution $\MeanPtSq$.

A low-$\pt$ cut-off on the $\jpsi$ candidates is applied due to the observation of a $\jpsi$ excess for $\pt<0.3$~\GeVc at forward rapidity in peripheral $\PbPb$ collisions at $\snn=2.76$~TeV~\cite{Adam:2015gba}, which is found to originate from coherent photo-production. Since this production mechanism is not normally included in hadro-production models, the low-$\pt$ interval is excluded for enabling comparisons with theoretical calculations. At midrapidity, mainly due to a better momentum resolution, nearly all of the coherent yield is contained in the range of reconstructed $\pt<0.15$~\GeVc, as shown by the ALICE measurements of $\jpsi$ photo-production in ultra-peripheral collisions~\cite{Abbas:2013oua}. A small component of incoherently photo-produced $\jpsi$ is still present in the range $\pt<1$~\GeVc, but for the centrality intervals considered in this work it is negligible. Thus, in the following, unless otherwise specified, all the results refer to $\jpsi$ with $\pt$ larger than $0.15$~\GeVc.

The inclusive $\jpsi$ nuclear modification factor is computed for a given centrality class as
\begin{equation}
\RAA = \frac{\text{d}^2 N/\text{d}y\text{d}\pT} {\langle \TAA \rangle \text{d}^2\sigma_{\text{pp}} /\der y \der \pT},
\label{eq:Raa}
\end{equation}
where $\text{d}^2 N/\text{d}y\text{d}p_{\text{T}}$ is the inclusive $\jpsi$ yield defined in Equation~\ref{eq:yield}, the $\langle \TAA \rangle$ is the average nuclear overlap function corresponding to the considered centrality class and $\text{d}^2\sigma_{\pp} /\text{d}y\text{d} \pt$ is the inclusive $\jpsi$ cross section measured by ALICE in $\pp$ collisions at $\s = 5.02$~TeV~\cite{Acharya:2019lkw}. The values used for the nuclear overlap function are shown in Table~\ref{tab:centrality}
and are obtained from~\cite{centralityNote}.

\begin{table}
  \label{tab:Glauber}
  \caption{Average number of participant nucleons $\AvgNpart$ and average nuclear overlap function $\langle \TAA \rangle$ for the centrality classes used in this analysis. 
   The values are derived from~\cite{centralityNote}.}\label{tab:centrality}
  \centering
   \begin{tabular}{ c | c  c }
    Centrality ($\%$) & $\AvgNpart$ &  $\AvgTAA$ (mb$^{-1}$)\\
    \hline
    
    $0$--$10$  & $357.3 \pm 0.8$ & $23.26 \pm 0.17$  \\
    $0$--$20$  & $309.7 \pm 0.9$ & $18.83 \pm 0.14$ \\
    $10$--$20$ & $262.0 \pm 1.2$  & $14.40 \pm 0.13$ \\
    $20$--$40$ & $159.4 \pm 1.3$  & $6.97 \pm 0.09$ \\
    $40$--$60$ & $70.7 \pm 0.9$ & $2.05 \pm 0.04$ \\
    $40$--$90$ & $39.0 \pm 0.7$ & $1.00 \pm 0.03$ \\
    $60$--$90$ & $17.9 \pm 0.3$ & $0.31 \pm 0.01$ \\
    \hline

    $0$--$90$ & $125.9 \pm 1.0$ & $6.28 \pm 0.12$\\
    \hline
  \end{tabular}
\end{table}

\section{Systematic uncertainties}

The systematic uncertainties on the measured $\jpsi$ yields, $\MeanPt$, and $\MeanPtSq$ originate from uncertainties on tracking, electron identification, signal extraction procedure, the kinematics used in the MC simulation for estimating the $\AccEff$ corrections, and the $\jpsi$ decay branching ratio. For the $\RAA$ and the $\raa$, the uncertainties on the $\jpsi$ cross section measurement in $\pp$ collisions~\cite{Acharya:2019lkw} and (only in the case of the $\RAA$) on the nuclear overlap function $\AvgTAA$~\cite{centralityNote} need to be considered in addition. A summary of the uncertainties on the $\pt$-integrated and $\pt$-differential yields is given in Table~\ref{tab:Systematics}.

The systematic uncertainty on the tracking of the candidate electrons is mainly due to uncertainties on the ITS-TPC track matching and on the track reconstruction efficiency in both the ITS and the TPC. These uncertainties, mainly due to differences in the reconstruction efficiency between data and MC, are estimated by varying the main track selection criteria and repeating the whole analysis chain. All variations which provide a corrected yield that deviates from the yield obtained with the standard selection criteria by more than one standard deviation are considered~\cite{Barlow}. The tracking uncertainty is then obtained as the root-mean-square of the distribution of all the valid variations, while the distribution mean is used as the central value. For the $\jpsi$ yields, this uncertainty ranges between $2$\% and $7$\% as a function of centrality (integrated over $\pt$) and between $4$\% and $9$\% as a function of transverse momentum. The tracking systematic uncertainty on the $\MeanPt$ and $\MeanPtSq$ are smaller than those for the corrected yields and detailed in Table~\ref{tab:Systematics}.

Uncertainties on the electron identification are due to the TPC electron PID response and the hadron rejection. A data-driven procedure is used to improve the matching between data and simulation for the electron selection by employing a pure sample of electrons from tagged photon conversions in the detector material. The residual mismatches are estimated by varying all the PID selection criteria following a similar procedure as for the tracking systematic uncertainty. The extracted uncertainty on the $\jpsi$ yields ranges between $1$\% and $6$\%, depending on the centrality and transverse momentum interval.

\begin{table}[t]
\begin{center}
\caption{Systematic uncertainties on the $\pt$-integrated and on $\pt$-differential $\jpsi$ yields for different centrality intervals. Only the ranges of uncertainty are quoted over the considered centrality intervals. The individual contributions and the total uncertainties are given as percent values.}
\label{tab:Systematics}
\begin{tabular}{c | c  c  c  c}

Source             & $\pt$-integrated & $\pt$-differential & $\MeanPt$ & $\raa$  \\
\hline
Tracking           &  2--7             & 4--9                &   2--4      &   3--6         \\ 
PID                &  3--6             & 1--6                &   1--2      &   2--4        \\
Signal extraction  &  2--7             & 5--7                &   1--2      &   2--3       \\
MC input           &  2                & 1--2                &   n.a.      &   n.a.       \\
\hline 
$\AvgTAA$               &  2--5             & 2--5                &    n.a.    &     n.a.       \\
pp reference       &  7                & 9--12               &     3      &     5       \\
\hline
\end{tabular}    
\end{center}
\end{table}

The uncertainty from the signal extraction procedure includes two components, one due to the $\jpsi$ signal shape and one due to the residual correlated dielectron background in the $\jpsi$ mass region. In order to estimate the uncertainty on the signal shape, the corrected $\jpsi$ yields are estimated using variations of the standard signal counting mass region, 2.92-3.16~\GeVcsq. For this we used three additional values of the lower mass limit, between 2.92 and 2.80~\GeVcsq and two additional values for the upper mass limit, namely 3.12 and 3.20~\GeVcsq. The correlated dielectron background in the invariant mass range used to extract the signal has generally a different shape compared to the combinatorial background. So matching the ME background in the sidebands of the same-event OS distribution may lead to a bias in the estimation of the raw yields. This is taken as a systematic uncertainty and is estimated by varying the mass ranges of the sidebands where the ME background is matched. By these variations the width of the sidebands was modified between 400 and 800~\MeVcsq. The total uncertainty on the signal extraction ranges between $2$\% to $7$\% as a function of centrality and between $5$\% and $7$\% as a function of $\pt$.

The acceptance and efficiency correction is $\pt$ dependent which makes correction factors averaged over large $\pt$ intervals sensitive to the $\jpsi$ $\pt$ spectrum used in the simulation. Since precise measurements of the $\jpsi$ transverse momentum spectra at midrapidity down to $\pt=0$~\GeVc are not available, the simulations used for corrections rely on the ALICE measurement at forward rapidity ($2.5<y<4.0$) in $\PbPb$ collisions at $\snn = 5.02$~TeV~\cite{Adam:2016rdg}. The measured spectrum including the statistical and systematic uncertainties is fitted using a power law function and the fit parameters are varied randomly within their allowed uncertainties taking into account their correlation matrix. The resulting uncertainty amounts to $2$\% for the $\pt$-integrated corrected yields and ranges between $1$\% to $2$\% in the considered $\pt$ intervals. 

Systematic uncertainties on the extraction of $\MeanPt$ and $\MeanPtSq$ are obtained by repeating the fit procedure with similar variations of tracking and PID selections as for the yield estimation. Since for this measurement the $\AccEff$ correction is applied for each dielectron pair using a fine-binned distribution of the correction factors, the systematic uncertainty due to the kinematics of the $\jpsi$ used in the MC simulation is negligible. In addition, the $\MeanPt$ and $\MeanPtSq$ are also extracted by directly fitting the corrected $\jpsi$ spectrum with a power law function and are found to be compatible to the values obtained from the fit with Equation~\ref{invMassFit}.

Systematic uncertainties on the tracking, PID and MC kinematics are considered to be partly correlated over both the centrality and the transverse momentum. The systematic uncertainties on signal extraction are considered as uncorrelated. The uncertainties on the nuclear overlap function are taken as uncorrelated over centrality and fully correlated over $\pt$ within a given centrality interval. The uncertainty on the $\pt$-integrated pp reference is considered to be fully correlated over centrality, while the uncertainties on the $\pt$-differential values are fully correlated over centrality and highly correlated over $\pt$.   

\section{Results and discussions}
\label{sec:Results}

\begin{figure}[t]
  \centering
  \includegraphics[width=0.48\textwidth]{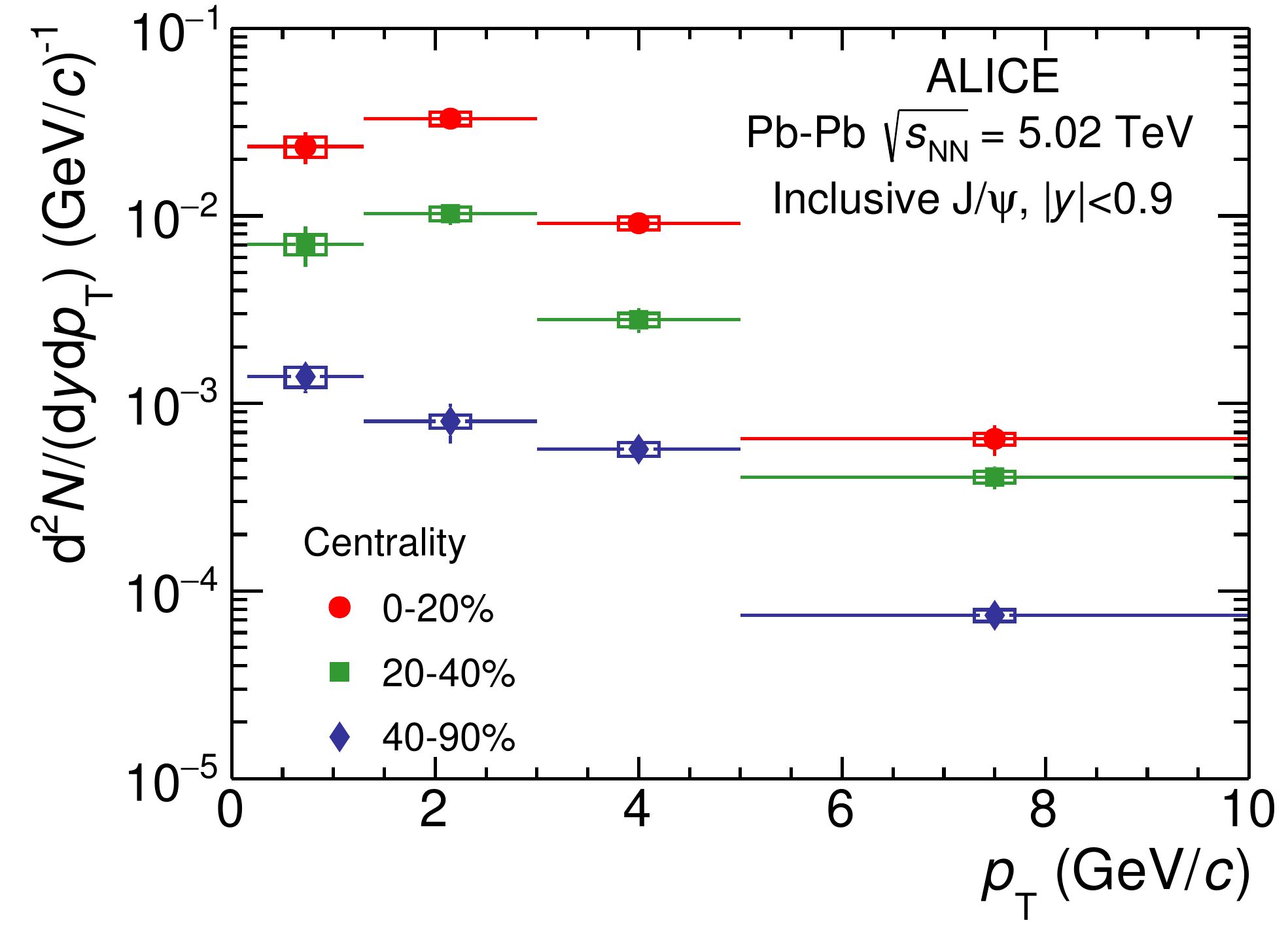}
  \includegraphics[width=0.48\textwidth]{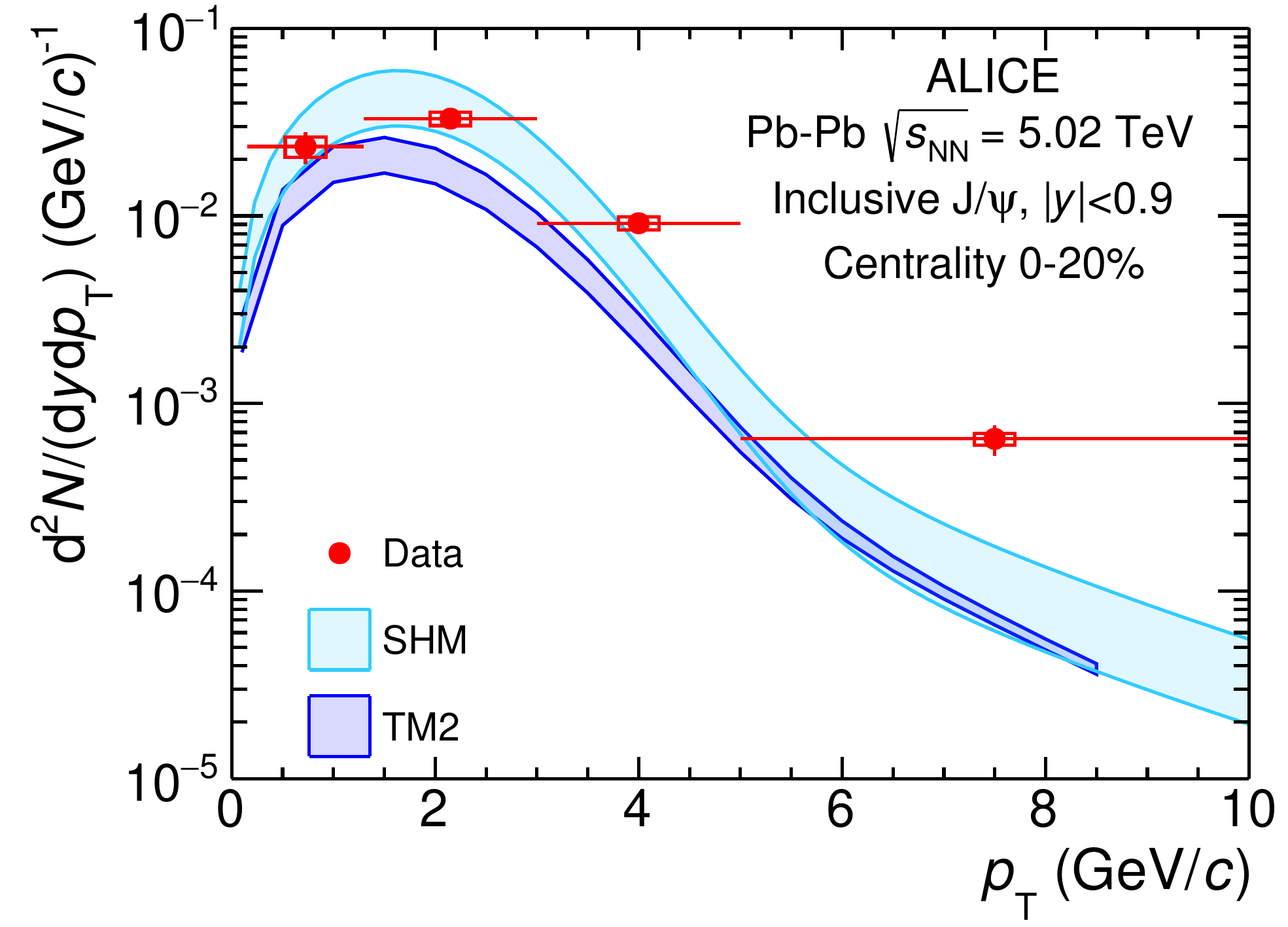}
\caption{(Color online) Left panel: Transverse momentum dependence of the $\jpsi$ production yields in $\PbPb$ collisions at $\snn = 5.02$~TeV at midrapidity in the centrality intervals $0$--$20$\%, $20$--$40$\%, and $40$--$90$\%. Right panel: Comparison of the $\pt$ distribution in the centrality interval $0$--$20$\% with models~\cite{Andronic:2019wva,Zhou:2014kka}.}\label{fig:JpsiPtSpectra}
\end{figure}

The inclusive $\jpsi$ $\pt$-differential yields evaluated using Equation~\ref{eq:yield} are shown in the left panel of Figure~\ref{fig:JpsiPtSpectra} (left) for the $0$--$20$\%, $20$--$40$\%, and $40$--$90$\% centrality intervals. The vertical error bars indicate statistical uncertainties while the systematic uncertainties, independently of their degree of correlation, are shown as boxes around the data points. The horizontal error bars show the evaluated $\pt$-range with the data point placed in the center.

The experimental results are compared with different phenomenological models of the charmonium production in relativistic heavy-ion collisions, i.e.\ the statistical hadronization model (SHM) by Andronic et al. ~\cite{Andronic:2007bi}, the comover interaction model (CIM) by Ferreiro ~\cite{Capella:1996va, Ferreiro:2012rq} and two different microscopic transport models, by Zhao et al. (TM1)~\cite{Zhao:2007hh} and by Zhou et al.(TM2)~\cite{Zhou:2014kka}.

In the SHM, all heavy quarks are produced during the initial hard partonic interactions followed by their thermalization in the QGP and the subsequent formation of bound states at the phase boundary according to their thermal weights. The $\pt$-integrated charmed-hadron yields depend only on the total $\ccBar$ cross section in heavy-ion collisions and on the chemical freeze-out parameters, which are determined by fitting measured light-flavored hadron yields. In addition to the high-density core part in the QGP, a corona contribution is added for the case that the nuclear density decreases below $10$\% of its maximal value, where no QGP is assumed and the number of $\jpsi$ is calculated from yields in $\pp$ collisions scaled by the number of binary nucleon--nucleon collision. A recent update of the SHM~\cite{Andronic:2019wva} uses a MUSIC (3+1)D~\cite{Schenke:2010nt} hydrodynamical simulation to extract the transverse flow velocity and the radial velocity profile of the freeze-out hyper-surface, such that the $\jpsi$ $\pt$ can be extracted from a blast-wave parameterization which follows a Hubble-like expansion~\cite{Florkowski_book}.

The CIM~\cite{Ferreiro:2012rq} was developed specifically for the description of charmonium suppression in heavy-ion collisions via its interactions with a comoving medium, either hadronic or partonic. The hot medium effects are modeled using a rate equation which contains a loss term for charmonium dissociation, and a gain term for (re)generation. In this model, the charmonium dissociation rate depends on the density of comovers, obtained from experimental measurements and on the charmonium dissociation cross section which is an energy-independent parameter of the model, fixed from fits to low energy data. Charmonium dissociation is balanced by the (re)generation component which depends on the primordial charm-quark cross section.

Both microscopic transport models considered here, TM1~\cite{Zhao:2007hh} and TM2~\cite{Zhou:2014kka}, solve the Boltzmann equation for charmonia ($\jpsi$, $\chi_c$ and $\psi'$) with dissociation and recombination terms. Each model considers the fireball evolution using implementations of ideal hydrodynamics which include both the deconfined and the hadronic phase separated by a first order phase transition. The dissociation rate in both models depends on the medium density and on a lattice-QCD-inspired charmonium binding energy (in TM1) or squared radius of the bound state (in TM2), all of them being functions of temperature. The (re)generation component is implemented using different approaches. In the TM1 calculations, it is based on the assumption that the charm quarks reach statistical equilibration after a relaxation time of about a few fm/$c$, while in the TM2 calculations the charm quarks are recombined using the same cross section as for the dissociation process and a thermalized distribution of charm quarks.

The primordial $\ccBar$ production cross section in $\PbPb$ collisions is a common input for all of the above mentioned models. There is so far no measurement of the $\ccBar$ cross section in $\PbPb$ or $\pp$ collisions at $\snn = 5.02$~TeV at midrapidity, which lead dominantly to the uncertainty of the models. The cross section in $\PbPb$ collisions is obtained from the total $\ccBar$ cross section in $\pp$ collisions $\text{d}\sigma_{\ccBar}/\text{d}y$ scaled by the average number of nucleon--nucleon collisions $\AvgNcoll$ in a given centrality class of $\PbPb$ collisions with additional CNM effects taken into account. For the rapidity interval used in this work, $|y|<0.9$, the value of $\der \sigma_{\ccBar}/\der y$ estimated for MB $\PbPb$ collisions is $0.53 \pm 0.10$~mb for the SHM, $0.76 \pm 0.13$~mb for TM1, $0.78 \pm 0.09$~mb for TM2 and $0.56 \pm 0.11$~mb for CIM.

The right panel of Figure~\ref{fig:JpsiPtSpectra} shows a comparison of the inclusive $\jpsi$ transverse momentum spectrum in the $20$\% most central $\PbPb$ collisions to calculations from the SHM and TM2 models. The bands indicate model uncertainties mainly due to the assumptions on the $\text{d}\sigma_{\ccBar}/\text{d}y$. Good agreement between data and the SHM predictions is observed in the low-$\pt$ region, while for $\pt \gtrsim 5$~\GeVc the calculations underestimate the data. The TM2 calculations underestimate the measured yields over the measured $\pt$ range. 

\begin{figure}[t]
  \centering
  \includegraphics[width=0.48\textwidth]{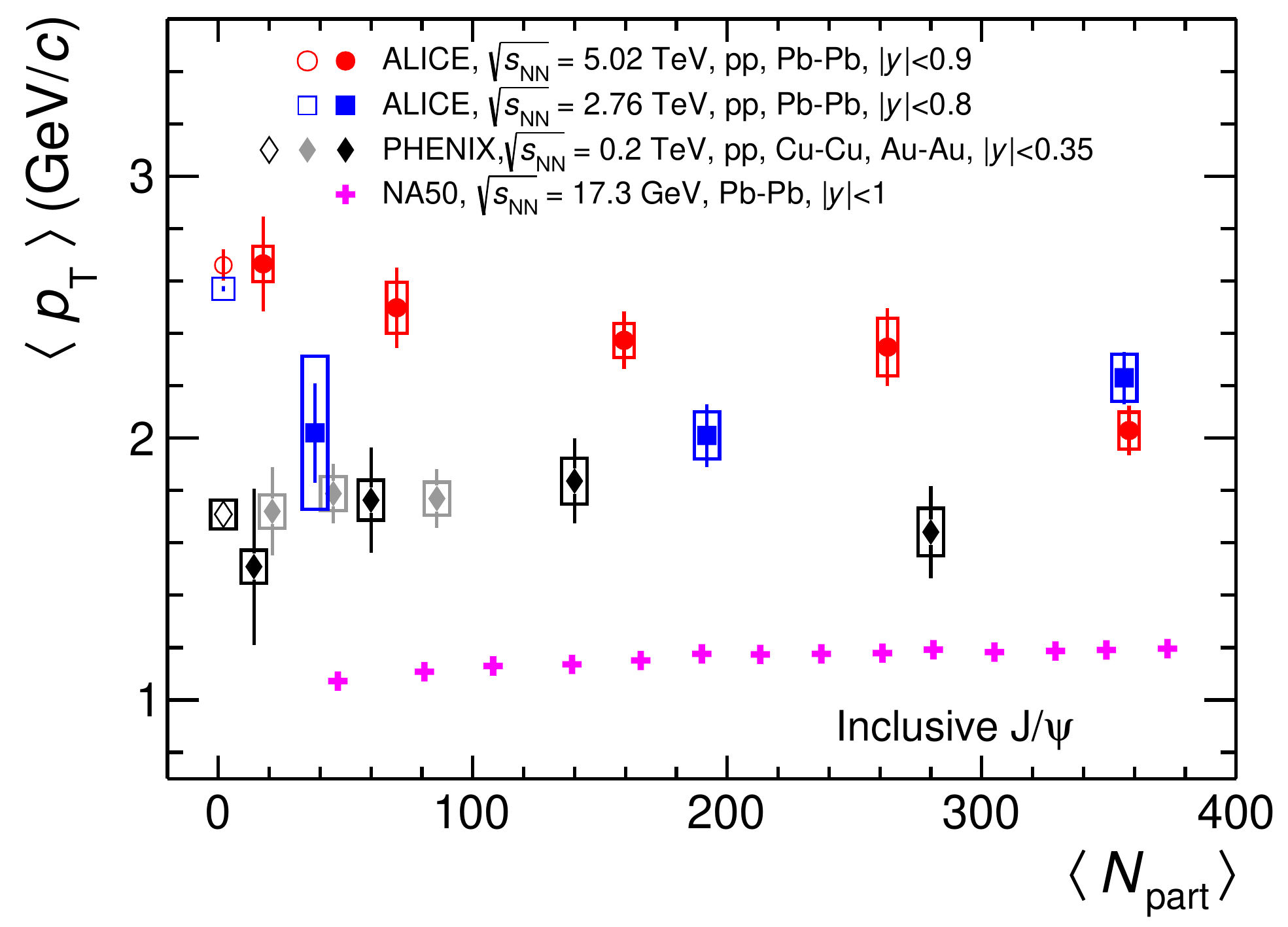}
  \includegraphics[width=0.48\textwidth]{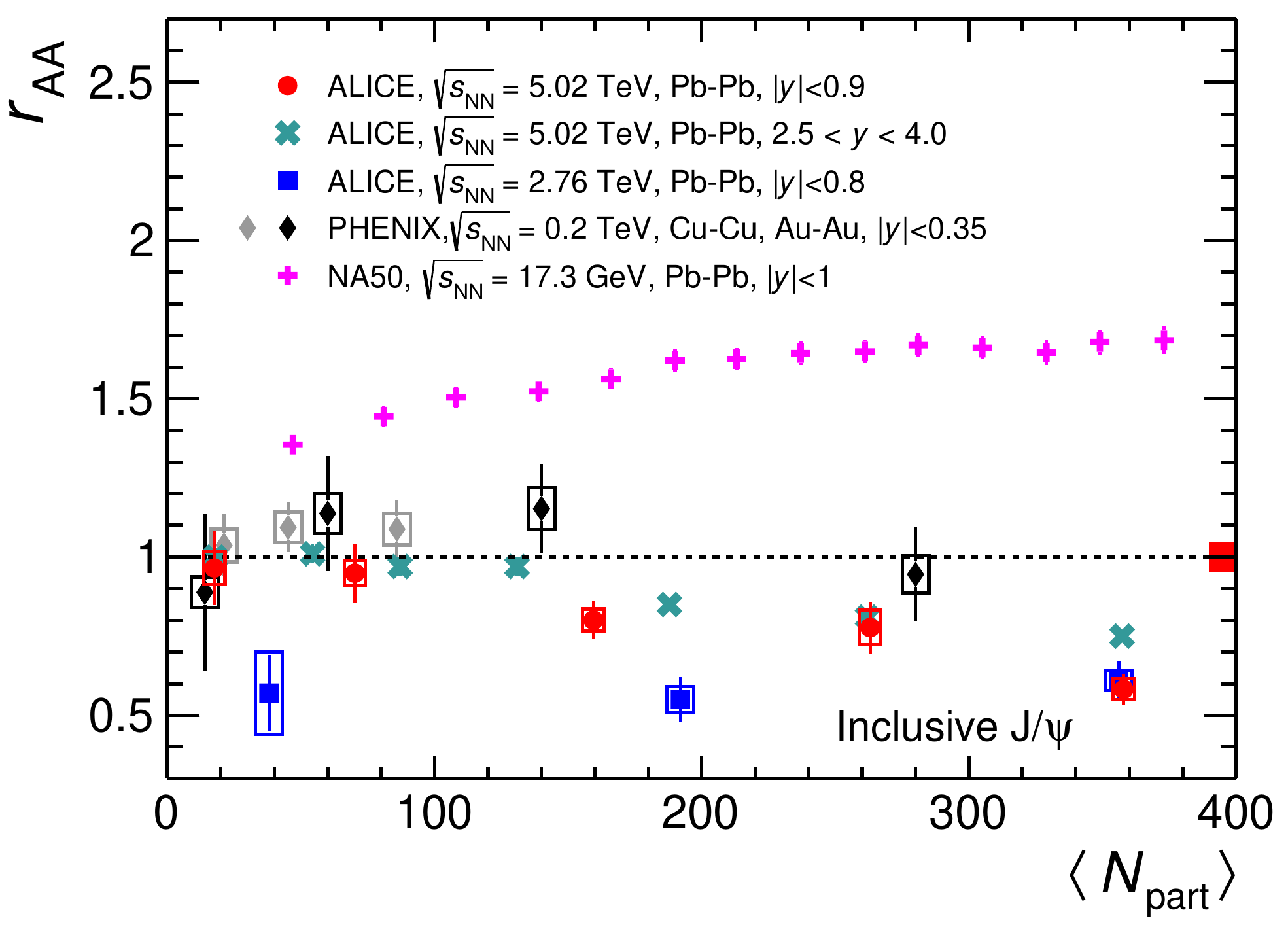}
  \caption{(Color online) $\jpsi$ $\MeanPt$ (left) and $\raa$ (right) at midrapidity as a function of the mean number of participant nucleons $\AvgNpart$. The ALICE measurements at $\snn = 5.02$~TeV are compared with previous results in $\pp$ and $\PbPb$ collisions at $2.76$~TeV~\cite{Adam:2015rba}, $\PbPb$ collisions at $5.02$ at forward rapidity~\cite{Acharya:2019iur}, and with those at lower collision energies at SPS~\cite{Abreu:2000xe} and RHIC~\cite{Adare:2011vq, Adare:2008sh, Adare:2006ns}. The red box around unity at $\Npart \approx 400$ in the right panel indicates the correlated uncertainty of the ALICE data points due to the $\MeanPtSq$ in $\pp$ collisions.}\label{Fig:MPtEnergy}
\end{figure}

\begin{figure}[t]
  \centering
  \includegraphics[width=0.48\textwidth]{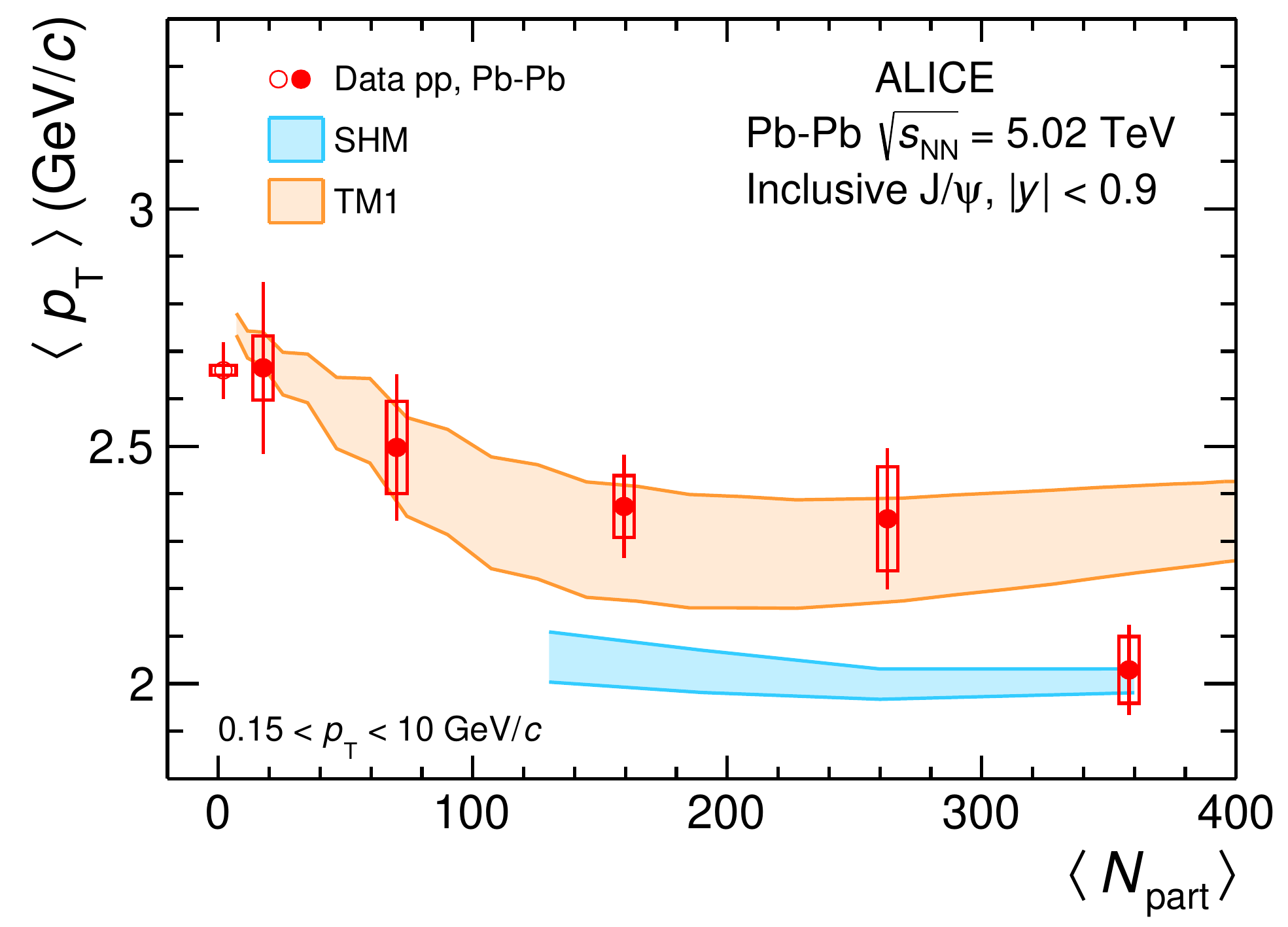}
  \includegraphics[width=0.48\textwidth]{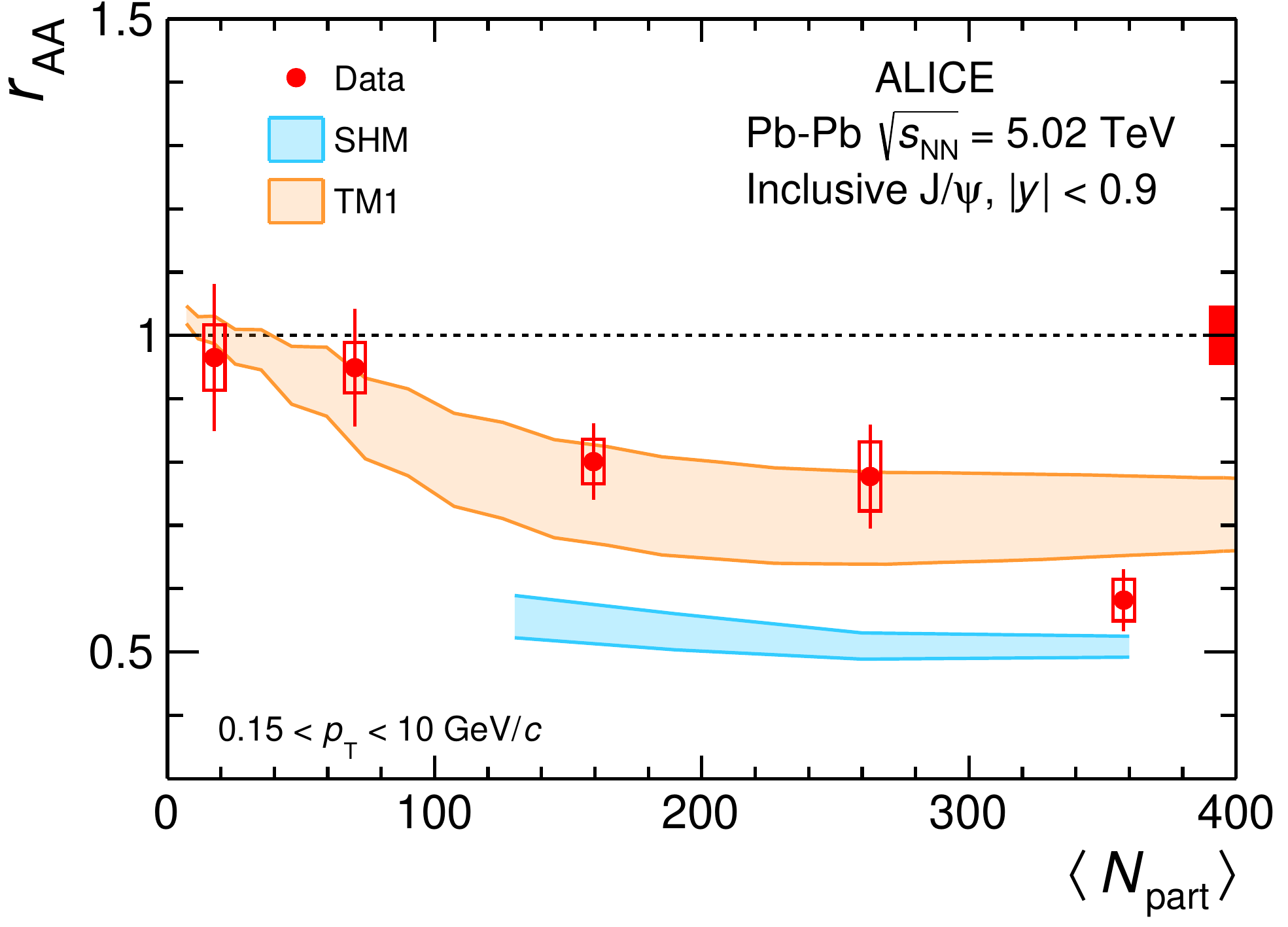}
  \caption{(Color online) Inclusive $\jpsi$ $\MeanPt$ (left) and $\raa$ (right) in $\pp$~\cite{Acharya:2019lkw} and $\PbPb$ collisions at $\snn = 5.02$~TeV at midrapidity as a function of the mean number of participating nucleons. The ALICE results are compared with calculations from the transport models~\cite{RappTransport1, RappTransport2} and the SHM~\cite{Andronic:2019wva}. The colored bands represent model uncertainties. As in Figure~\ref{Fig:MPtEnergy}, the red box around unity at $\Npart \approx 400$ in the right panel indicates the correlated uncertainty of the ALICE data points due to the $\MeanPtSq$ in $\pp$ collisions.}\label{Fig:MPtModels}
\end{figure}

In order to facilitate the comparison of the $\jpsi$ $\pt$ spectra obtained in this work with other measurements or theory calculations, the $\jpsi$ $\MeanPt$ and $\MeanPtSq$ are extracted in several centrality intervals, using the method described in Sec.~\ref{sec:Analysis}. The left panel of Figure~\ref{Fig:MPtEnergy} shows the $\jpsi$ $\MeanPt$ dependence on the mean number of participant nucleons $\AvgNpart$. The $\MeanPt$ in $\PbPb$ collisions at $\snn = 5.02$~TeV shows a monotonic decrease from the most peripheral collisions, where it is compatible to the measurement in $\pp$ collisions at $\s=5.02$~TeV, to the most central collisions, which hints towards a strong contribution from (re)combination processes. This trend is not clearly visible for the measurement at $\snn = 2.76$~TeV, which suffered from large statistical and systematic uncertainties.

The $\raa$ ratio, defined as $\MeanPtSq_{\text{PbPb}} / \MeanPtSq_{\mathrm{pp}}$, which is shown in the right panel of Figure~\ref{Fig:MPtEnergy}, is a measure of the broadness of the $\pt$ spectra in heavy-ion collisions relative to $\pp$ collisions at the same energy. A strong decrease of the $\raa$ is observed in $\PbPb$ collisions at $\snn = 5.02$~TeV between peripheral, where it is consistent with unity, and central collisions where $\raa$ reaches a value of $0.6$ at midrapidity and $0.75$ at forward rapidity~\cite{Acharya:2019iur}. When comparing with measurements at lower energies from RHIC~\cite{Adare:2011vq, Adare:2008sh, Adare:2006ns} and SPS~\cite{Abreu:2000xe}, a very different picture emerges. While the RHIC measurements for both $\MeanPt$ and $\raa$ are compatible with a constant trend as a function of $\AvgNpart$~\cite{Zhou:2014kka}, the SPS results show a monotonic increase of both $\MeanPt$ and $\raa$ as a function of collision centrality which, at this energy, can be explained by a broadening of the $\pt$ distribution due to the Cronin effect~\cite{Cronin:1974zm}. 

The results for the $\MeanPt$ and $\raa$ in $\PbPb$ collisions at $\snn = 5.02$~TeV are compared with model calculations in Figure~\ref{Fig:MPtModels}. The statistical hadronization model agrees with the data only for the most central collisions but underestimates the measurements for more peripheral collisions. A good description of the centrality trend is obtained with the transport model TM1 calculation, which includes a detailed implementation of the fireball evolution, with the exception of most central collisions where the model overestimates both the $\jpsi$ $\MeanPt$ and $\raa$. 

\begin{figure}[t] 
  \begin{center}
    \includegraphics[width = 7.7 cm]{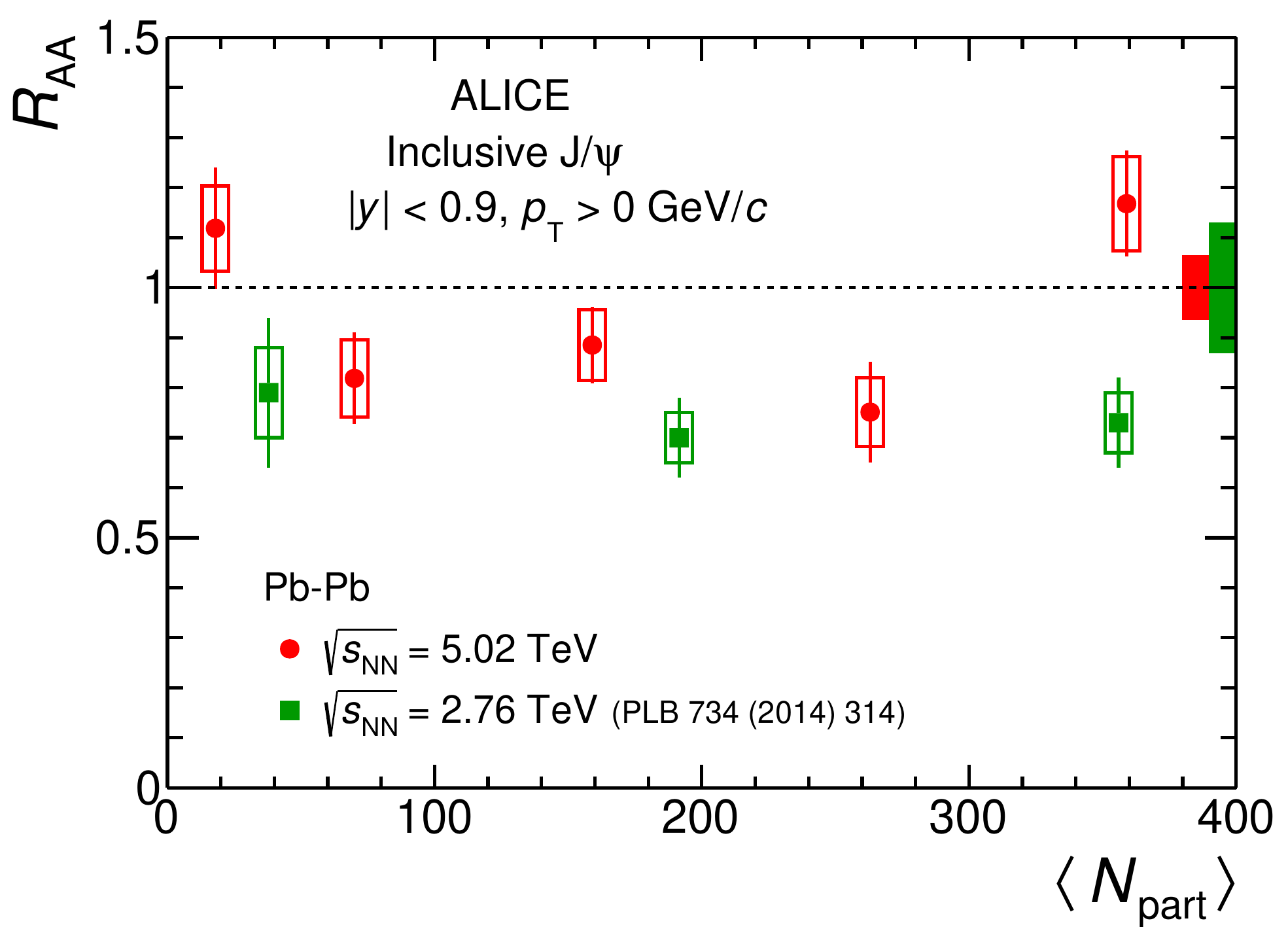}
    \includegraphics[width = 7.7 cm]{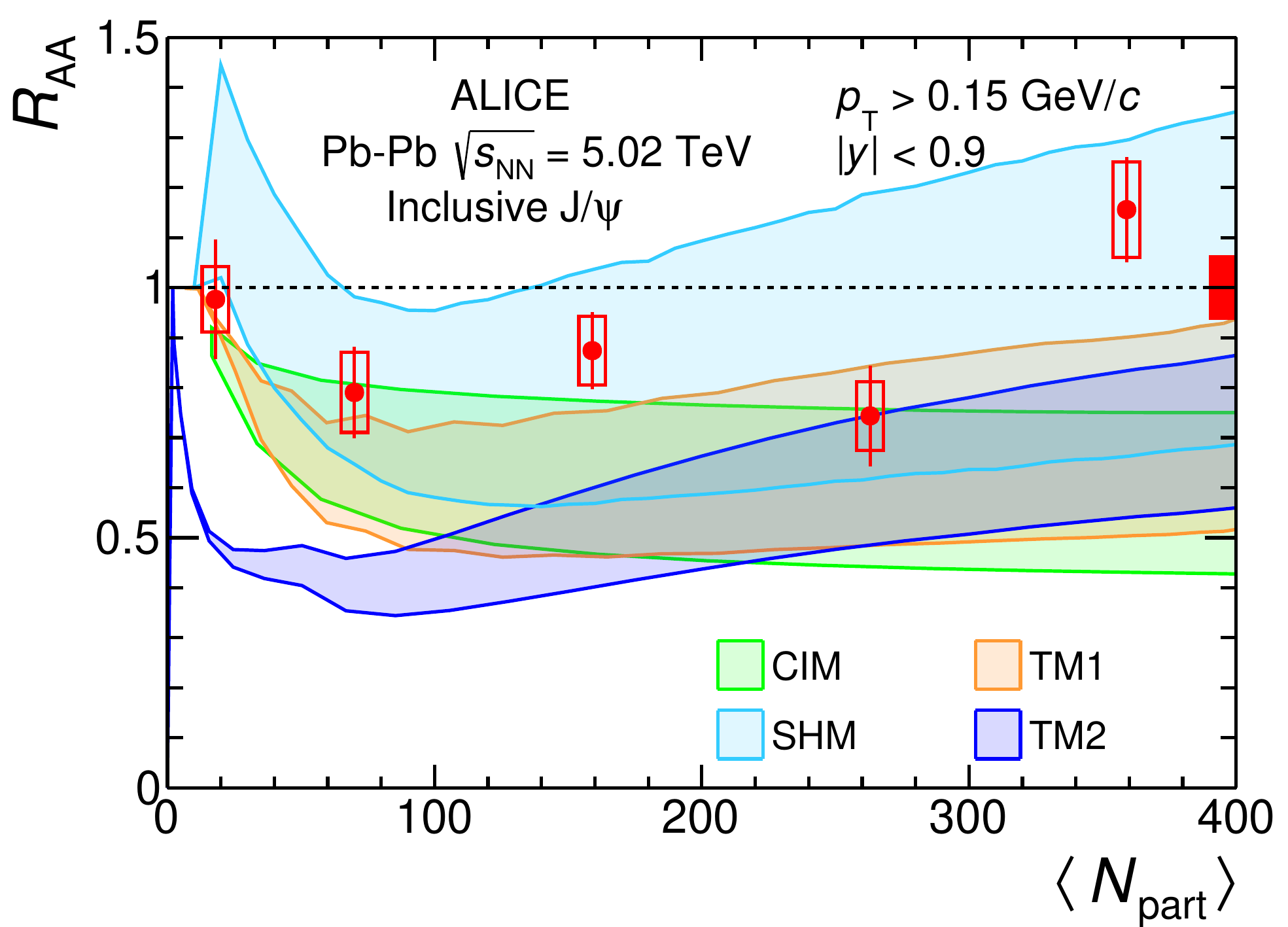}
    \caption{(Color online) Inclusive $\jpsi$ nuclear modification factor at midrapidity, integrated over $\pt$, as a function of $\langle\npart\rangle$ in $\PbPb$ collisions at $\snn = 5.02$~TeV compared with results at $\snn = 2.76$~TeV~\cite{Abelev:2013ila} (left panel) and with calculations from the CIM~\cite{Ferreiro:2012rq}, SHM~\cite{Andronic:2019wva}, TM1~\cite{RappTransport2} and TM2~\cite{Zhou:2014kka} models (right panel). The yields in the left panel are shown without the low-$\pt$ cut-off in order to be able to compare with the lower energy data which are obtained for $\pt>0$. The calculations are shown as bands indicating the model uncertainties. Boxes around unity at $\Npart \approx 400$ in both panels indicate the correlated uncertainty of the data points due to the cross section in $\pp$ collisions.}
\label{fig:raaInclusive}
\end{center}
\end{figure}

The $\pt$-integrated nuclear modification factor for inclusive $\jpsi$ in $\PbPb$ collisions at $\snn = 5.02$~TeV obtained using Equation~\ref{eq:Raa} is shown in the left panel of Figure~\ref{fig:raaInclusive} as a function of the mean number of participants and compared with a measurement at $\snn=2.76$~TeV~\cite{Abelev:2013ila}. The boxes shown around unity indicate the correlated systematic uncertainties and include the uncertainties on the $\pp$ reference. Besides the most central collisions where there is a hint of an increase of the $\RAA$ with collision energy, the results at the two energies are compatible within uncertainties. A comparison of the experimental results at $\snn = 5.02$~TeV with calculations based on the models described before is shown in the right panel of Figure~\ref{fig:raaInclusive}. The calculations are shown as bands that indicate model uncertainties, dominated by the uncertainties on the $\ccBar$ cross section and on the CNM effects. The SHM calculation shows a good agreement with the data over the entire centrality range. CIM, TM1 and TM2 calculations underestimate the experimental results towards the data points corresponding to the most central collisions despite the fact that the total $\ccBar$ cross section assumed in TM1 and TM2 is significantly larger compared to the SHM and the CIM. The large model uncertainties do not allow a conclusion to be made on the phenomenology of charmonium production in nuclear collisions. This emphasizes the importance of a precise measurement of the total $\ccBar$ cross section, but also the need of using consistent model inputs, including the total $\ccBar$ cross section, the $\pp$ reference $\jpsi$ cross section and CNM effects.  

\begin{figure}[t] 
  \begin{center}
    \includegraphics[width = 7.7 cm]{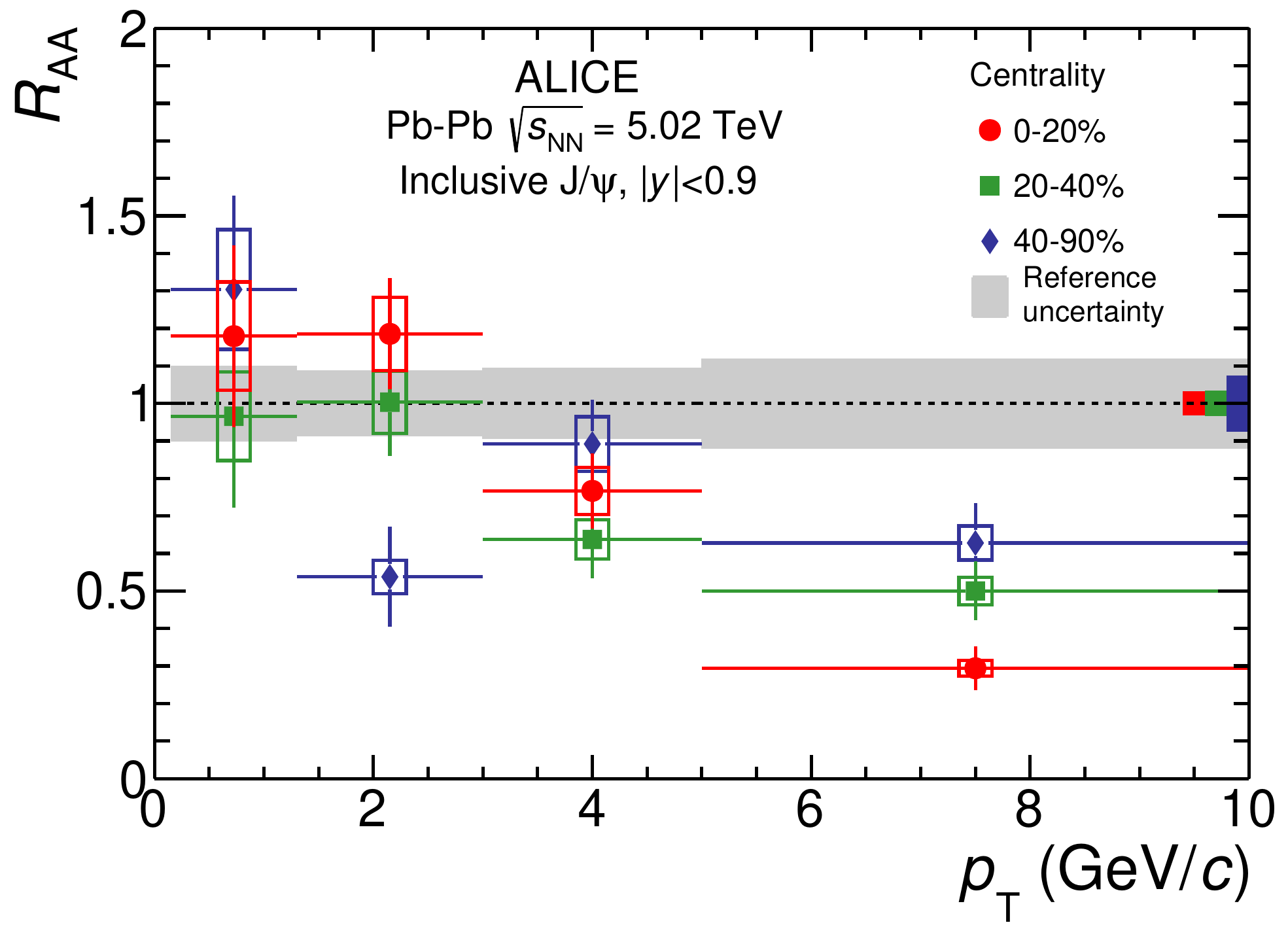}
    \includegraphics[width = 7.7 cm]{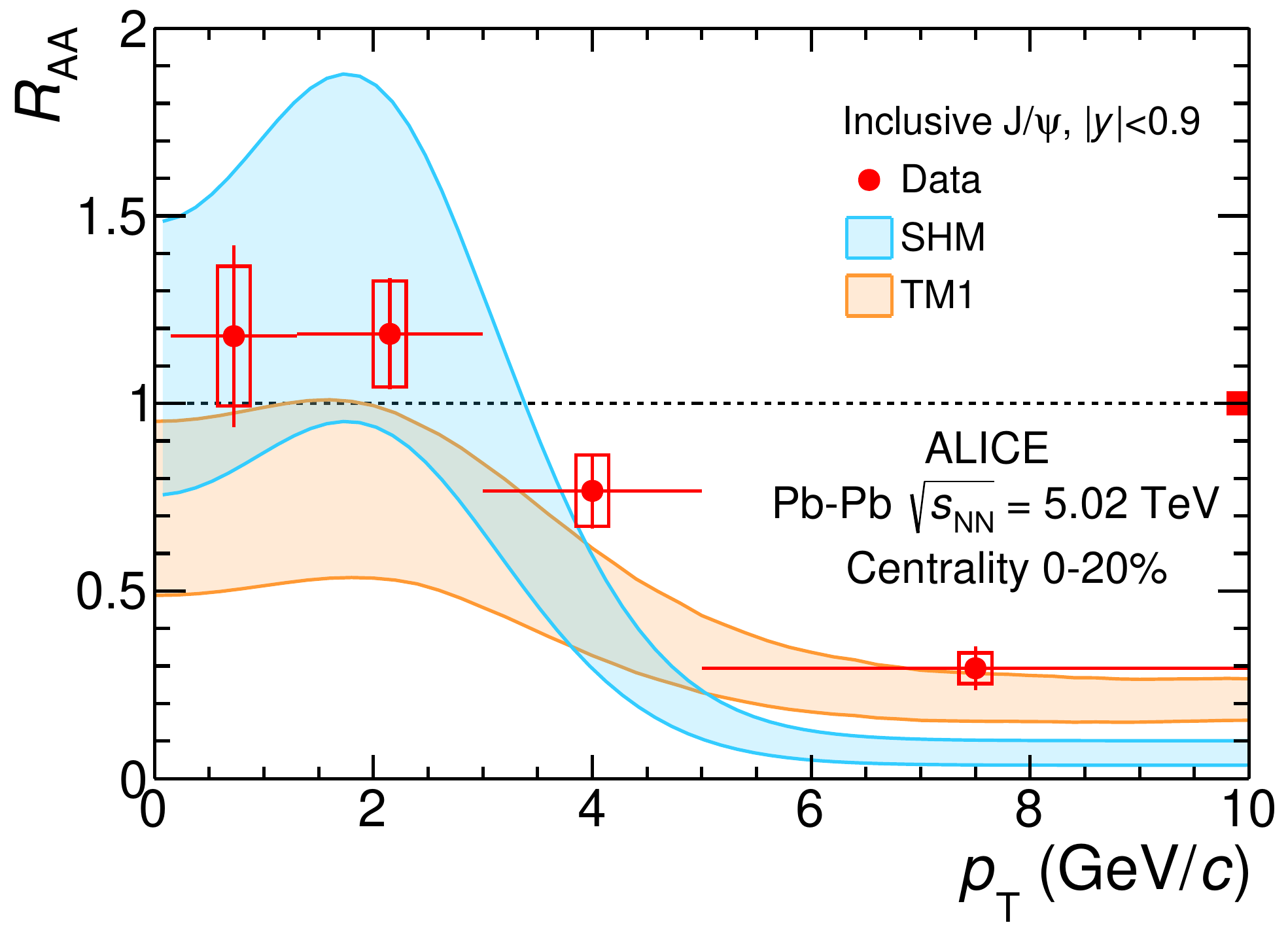}
    \caption{(Color online) Inclusive $\jpsi$ $\Raa$ at midrapidity in $\PbPb$ collisions at $\snn = 5.02$~TeV as a function of $\pt$ for different centrality intervals (left) and compared with model calculations~\cite{Andronic:2019wva, RappTransport2} for the centrality interval $0$--$20$\% (right).}
    \label{fig:raavspt}
  \end{center}
\end{figure}

The inclusive $\jpsi$ nuclear modification factor in $\PbPb$ collisions as a function of $\pt$ is shown in the left panel of Figure~\ref{fig:raavspt} for the centrality intervals $0$--$20$\%, $20$--$40$\% and $40$--$90$\%. The systematic uncertainties shown as boxes around the data points include the systematic uncertainties from the $\PbPb$ analysis while the uncertainties from the $\pp$ reference, correlated over centrality, are shown as the gray band around unity. The colored boxes at high $\pt$ around unity indicate the correlated uncertainties due to the $\AvgTAA$ values used for the $\RAA$ calculation. These results are compatible with binary scaling for $\pt<3$~\GeVc, with the exception of the data point around $2$~\GeVc which shows a downward statistical fluctuation for $40$--$90$\% centrality, while the $\jpsi$ production is suppressed at higher $\pt$. With the current uncertainties it is difficult to extract a centrality trend except for the highest $\pt$ interval, $5$--$10$~\GeVc, where a stronger suppression is observed in the most central collisions relative to the more peripheral centrality intervals at a significance level of about 3$\sigma$. The results for the $20$\% most central collisions are compared with model calculations and shown in the right panel of Figure~\ref{fig:raavspt}. Both the SHM and TM1 models describe qualitatively the data. In these models, the increasing $\RAA$ towards low $\pt$ is a consequence of the dominant contribution of (re)generated $\jpsi$. At high $\pt$, the contribution from recombination decreases, and the $\jpsi$ production is suppressed due to color charges in the medium. The main $\jpsi$ sources at high $\pt$ are primordial production and feed-down from beauty decays. The SHM, where the $\jpsi$ at high $\pt$ are produced only in the corona, overestimates the degree of $\jpsi$ suppression.

\begin{figure}[t] 
  \begin{center}
    \includegraphics[width = 7.7 cm]{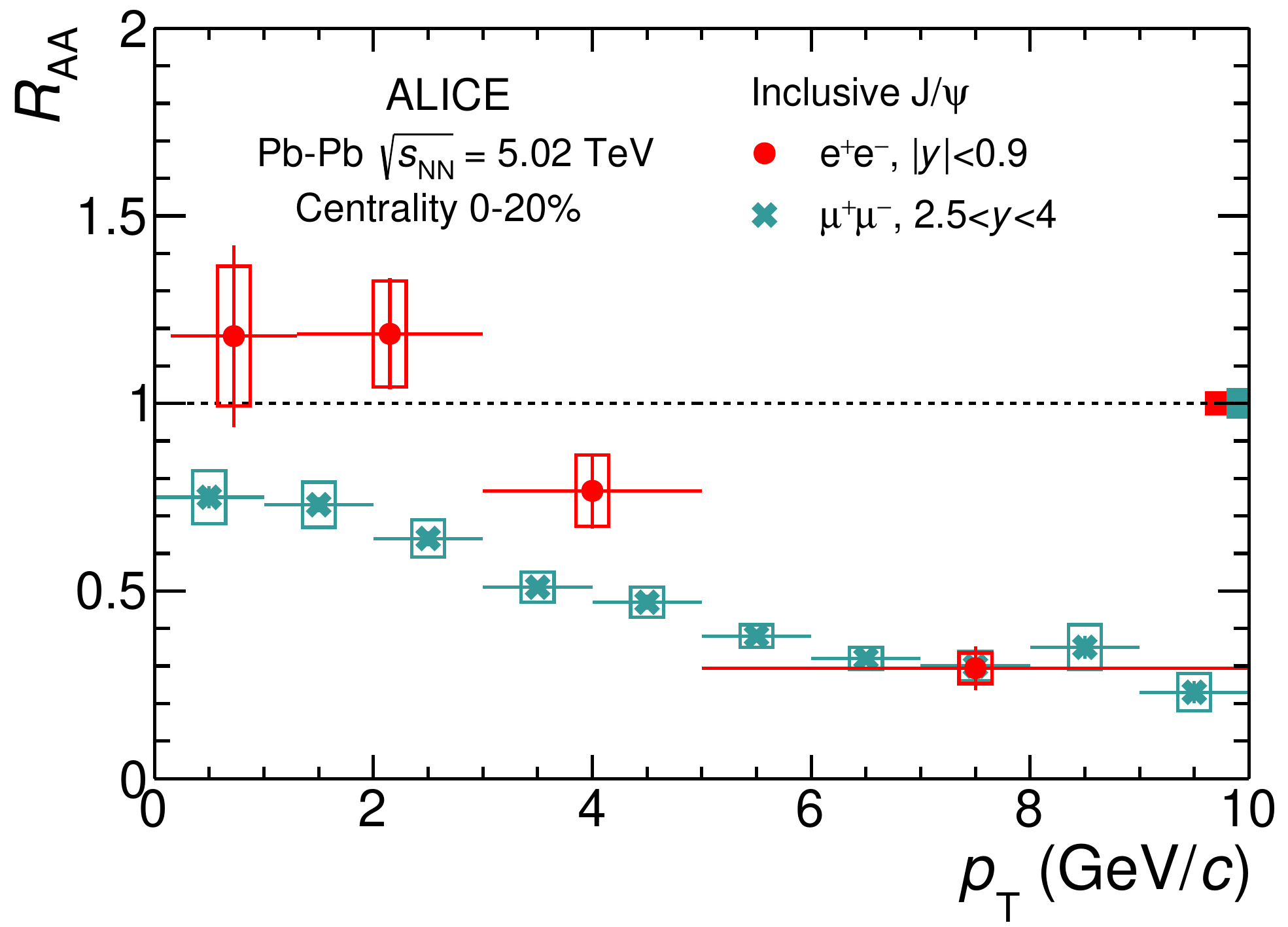}
    \includegraphics[width = 7.7 cm]{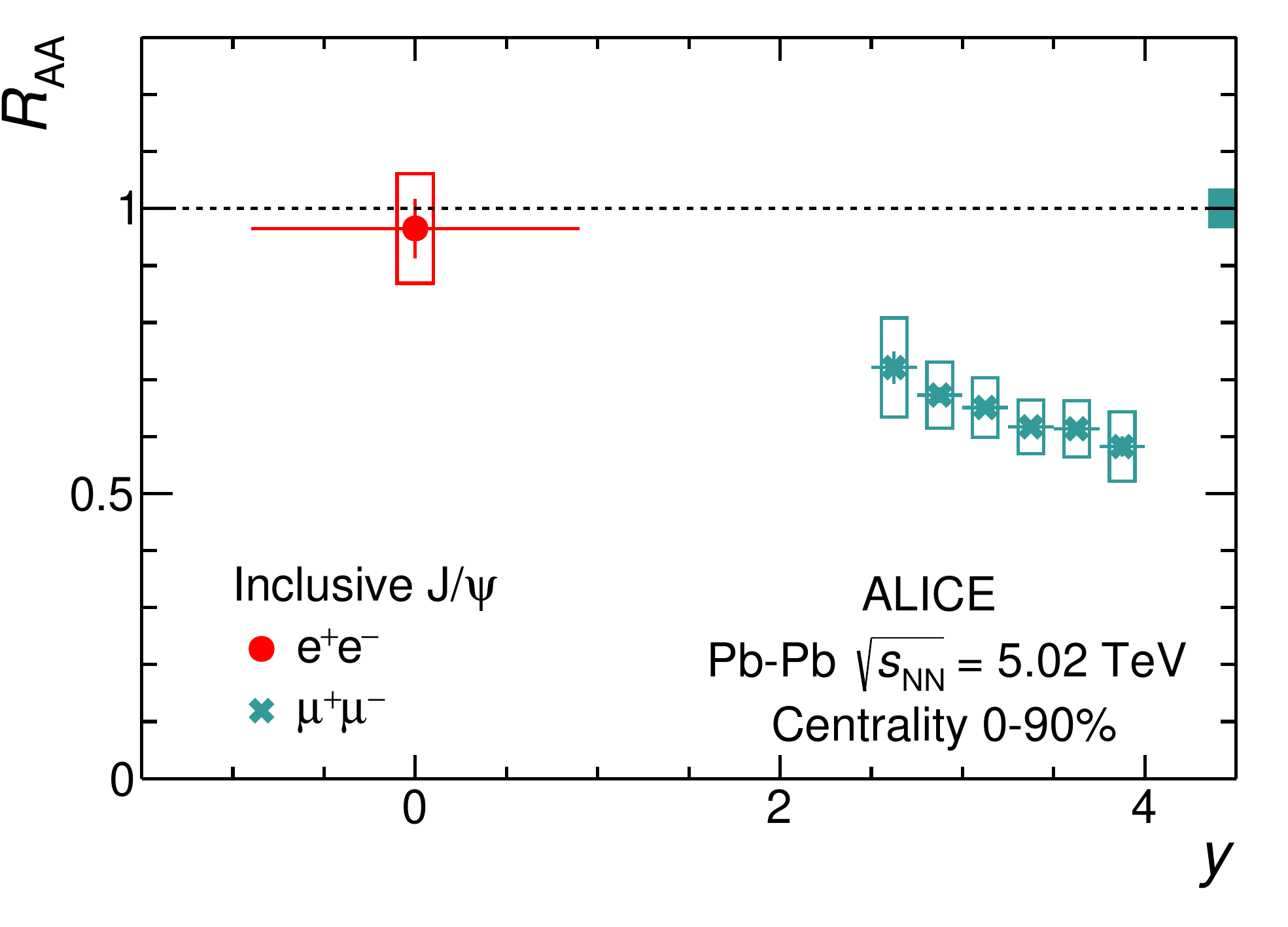}
    \caption{(Color online) Left: Inclusive $\jpsi$ $\Raa$ in the $20$\% most central $\PbPb$ collisions at $\snn = 5.02$~TeV as a function of $\pt$, at midrapidity and at forward rapidity~\cite{Adam:2016rdg,Acharya:2019iur}. Right: Rapidity dependence of the inclusive $\jpsi$ $\Raa$ in the centrality interval $0$--$90$\%. The error bars represent statistical uncertainties, while the boxes around the data points represent systematic uncertainties. The boxes around unity represent global uncertainties on the $\pp$ reference due to normalization and $\langle\TAA\rangle$. In the right panel, the correlated uncertainty of the point at midrapidity is included in the box around the data point.}\label{fig:raavsy}
  \end{center}
\end{figure}

Since the charm quark density, i.e.\ the $\ccBar$ cross section, is expected to decrease towards larger rapidity, the comparison to the forward-rapidity measurements is a valuable source of information. In the left panel of Figure~\ref{fig:raavsy}, the $\pt$ dependence of the $\jpsi$ $\Raa$ in the $20$\% most central $\PbPb$ collisions at midrapidity is compared with the ALICE results measured at forward rapidity ($2.5<y<4$)~\cite{Adam:2016rdg}. The boxes around the data points represent systematic uncertainties, while the boxes drawn around $\Raa=1$ show global uncertainties on the pp reference due to uncertainties on the beam luminosity and $\AvgTAA$. In the low-$\pt$ range ($\pt<5$~\GeVc) these data indicate larger $\Raa$ values at midrapidity compared to those at forward rapidity, with a combined statistical significance of nearly 4$\sigma$, compatible with expectations from a (re)generation scenario due to the larger primordial $\ccBar$ density at midrapidity. The rapidity dependence of the inclusive $\jpsi$ suppression, integrated over $\pt$, is shown in the right panel of Figure~\ref{fig:raavsy} for the $0$--$90$\% centrality interval. The value of the $\Raa$ at midrapidity is $0.97 \pm 0.05 (\text{stat.}) \pm 0.1 (\text{syst.})$, and a monothonic decrease is observed towards forward rapidity~\cite{Adam:2016rdg,Acharya:2019iur}. 

\section{Conclusions}
\label{sec:Conclusions}

The measurements of the inclusive $\jpsi$ yields and nuclear modification factors at midrapidity ($|y| < 0.9$) were performed in the dielectron decay channel in $\PbPb$ collisions at a center-of-mass energy $\snn = 5.02$~TeV using an integrated luminosity of $L_{\text{int}} \approx 10~\mu \text{b}^{-1}$ collected by the ALICE Collaboration. The results were presented as a function of transverse momentum in different collision centrality classes.

The $\jpsi$ transverse momentum dependent yields in central $\PbPb$ collisions are well reproduced in the low $\pt$ range by an updated SHM calculation~\cite{Andronic:2019wva} and underestimated for large $\pt$. The TM2~\cite{Zhou:2014kka} transport calculations underestimate the $\jpsi$ yields over the entire measured $\pt$ range. The $\jpsi$ $\MeanPt$ and $\MeanPtSq$ show a decrease from peripheral collisions, where they are similar to the values observed in pp collisions, towards most central collisions. This centrality-dependent behavior is qualitatively different compared to the observations at lower energies from RHIC and SPS and can be explained through the interplay between the (re)generation process, dominant at low $\pt$ for central events at the LHC, color screening, and CNM effects like gluon shadowing. A good description of the observed trends is provided by the TM1 calculations, while the SHM calculations agree with the data for central collisions only. 

The $\pt$-integrated nuclear modification factor as a function of the number of participant nucleons shows a moderate level of suppression in the range $50 < \AvgNpart < 300$, and indicates an increase towards central collisions. In the most peripheral collisions, our results are compatible with binary scaling of the $\jpsi$ production. The nuclear modification factor as a function of the transverse momentum shows a strong suppression, centrality dependent, for $\pt > 3~\GeVc$ but is compatible with unity or with a small enhancement at small $\pT$, suggestive of the large contribution from the (re)generation process. Furthermore, from these measurements we observe significantly larger values for $\Raa$ compared to the results at forward rapidity~\cite{Adam:2016rdg} for both the $\pt$-integrated values in the $0$--$90$\% centrality interval and for the $\pt$-differential $\Raa$ in the low $\pt$ region ($\pt<5$~\GeVc) in the centrality interval $0$--$20$\%. 

Consequently, these results strenghten the hypothesis that charmonium at low $\pt$ is produced predominantly via (re)generation in the late stages of the collision at the LHC. However, due to the remaining experimental and theoretical uncertainties, the exact phenomenology leading to these observations cannot be determined yet.

\newenvironment{acknowledgement}{\relax}{\relax}
\begin{acknowledgement}
\section*{Acknowledgements}

The ALICE Collaboration would like to thank all its engineers and technicians for their invaluable contributions to the construction of the experiment and the CERN accelerator teams for the outstanding performance of the LHC complex.
The ALICE Collaboration gratefully acknowledges the resources and support provided by all Grid centres and the Worldwide LHC Computing Grid (WLCG) collaboration.
The ALICE Collaboration acknowledges the following funding agencies for their support in building and running the ALICE detector:
A. I. Alikhanyan National Science Laboratory (Yerevan Physics Institute) Foundation (ANSL), State Committee of Science and World Federation of Scientists (WFS), Armenia;
Austrian Academy of Sciences, Austrian Science Fund (FWF): [M 2467-N36] and Nationalstiftung f\"{u}r Forschung, Technologie und Entwicklung, Austria;
Ministry of Communications and High Technologies, National Nuclear Research Center, Azerbaijan;
Conselho Nacional de Desenvolvimento Cient\'{\i}fico e Tecnol\'{o}gico (CNPq), Financiadora de Estudos e Projetos (Finep), Funda\c{c}\~{a}o de Amparo \`{a} Pesquisa do Estado de S\~{a}o Paulo (FAPESP) and Universidade Federal do Rio Grande do Sul (UFRGS), Brazil;
Ministry of Education of China (MOEC) , Ministry of Science \& Technology of China (MSTC) and National Natural Science Foundation of China (NSFC), China;
Ministry of Science and Education and Croatian Science Foundation, Croatia;
Centro de Aplicaciones Tecnol\'{o}gicas y Desarrollo Nuclear (CEADEN), Cubaenerg\'{\i}a, Cuba;
Ministry of Education, Youth and Sports of the Czech Republic, Czech Republic;
The Danish Council for Independent Research | Natural Sciences, the VILLUM FONDEN and Danish National Research Foundation (DNRF), Denmark;
Helsinki Institute of Physics (HIP), Finland;
Commissariat \`{a} l'Energie Atomique (CEA), Institut National de Physique Nucl\'{e}aire et de Physique des Particules (IN2P3) and Centre National de la Recherche Scientifique (CNRS) and R\'{e}gion des  Pays de la Loire, France;
Bundesministerium f\"{u}r Bildung und Forschung (BMBF) and GSI Helmholtzzentrum f\"{u}r Schwerionenforschung GmbH, Germany;
General Secretariat for Research and Technology, Ministry of Education, Research and Religions, Greece;
National Research, Development and Innovation Office, Hungary;
Department of Atomic Energy Government of India (DAE), Department of Science and Technology, Government of India (DST), University Grants Commission, Government of India (UGC) and Council of Scientific and Industrial Research (CSIR), India;
Indonesian Institute of Science, Indonesia;
Centro Fermi - Museo Storico della Fisica e Centro Studi e Ricerche Enrico Fermi and Istituto Nazionale di Fisica Nucleare (INFN), Italy;
Institute for Innovative Science and Technology , Nagasaki Institute of Applied Science (IIST), Japanese Ministry of Education, Culture, Sports, Science and Technology (MEXT) and Japan Society for the Promotion of Science (JSPS) KAKENHI, Japan;
Consejo Nacional de Ciencia (CONACYT) y Tecnolog\'{i}a, through Fondo de Cooperaci\'{o}n Internacional en Ciencia y Tecnolog\'{i}a (FONCICYT) and Direcci\'{o}n General de Asuntos del Personal Academico (DGAPA), Mexico;
Nederlandse Organisatie voor Wetenschappelijk Onderzoek (NWO), Netherlands;
The Research Council of Norway, Norway;
Commission on Science and Technology for Sustainable Development in the South (COMSATS), Pakistan;
Pontificia Universidad Cat\'{o}lica del Per\'{u}, Peru;
Ministry of Science and Higher Education and National Science Centre, Poland;
Korea Institute of Science and Technology Information and National Research Foundation of Korea (NRF), Republic of Korea;
Ministry of Education and Scientific Research, Institute of Atomic Physics and Ministry of Research and Innovation and Institute of Atomic Physics, Romania;
Joint Institute for Nuclear Research (JINR), Ministry of Education and Science of the Russian Federation, National Research Centre Kurchatov Institute, Russian Science Foundation and Russian Foundation for Basic Research, Russia;
Ministry of Education, Science, Research and Sport of the Slovak Republic, Slovakia;
National Research Foundation of South Africa, South Africa;
Swedish Research Council (VR) and Knut \& Alice Wallenberg Foundation (KAW), Sweden;
European Organization for Nuclear Research, Switzerland;
Suranaree University of Technology (SUT), National Science and Technology Development Agency (NSDTA) and Office of the Higher Education Commission under NRU project of Thailand, Thailand;
Turkish Atomic Energy Agency (TAEK), Turkey;
National Academy of  Sciences of Ukraine, Ukraine;
Science and Technology Facilities Council (STFC), United Kingdom;
National Science Foundation of the United States of America (NSF) and United States Department of Energy, Office of Nuclear Physics (DOE NP), United States of America.    
\end{acknowledgement}

\bibliographystyle{utphys}   
\bibliography{alicepreprint_jpsi}

\newpage
\appendix
\section{The ALICE Collaboration}
\label{app:collab}

\begingroup
\small
\begin{flushleft}
S.~Acharya\Irefn{org141}\And 
D.~Adamov\'{a}\Irefn{org94}\And 
A.~Adler\Irefn{org74}\And 
J.~Adolfsson\Irefn{org80}\And 
M.M.~Aggarwal\Irefn{org99}\And 
G.~Aglieri Rinella\Irefn{org33}\And 
M.~Agnello\Irefn{org30}\And 
N.~Agrawal\Irefn{org10}\textsuperscript{,}\Irefn{org53}\And 
Z.~Ahammed\Irefn{org141}\And 
S.~Ahmad\Irefn{org16}
% ...
\And
\renewcommand\labelenumi{\textsuperscript{\theenumi}~}

\section*{Affiliation notes}
\renewcommand\theenumi{\roman{enumi}}
\begin{Authlist}
\item \Adef{org*}Deceased
\item \Adef{orgI}Dipartimento DET del Politecnico di Torino, Turin, Italy
\item \Adef{orgII}M.V. Lomonosov Moscow State University, D.V. Skobeltsyn Institute of Nuclear, Physics, Moscow, Russia
\item \Adef{orgIII}Department of Applied Physics, Aligarh Muslim University, Aligarh, India
\item \Adef{orgIV}Institute of Theoretical Physics, University of Wroclaw, Poland
\end{Authlist}

\section*{Collaboration Institutes}
\renewcommand\theenumi{\arabic{enumi}~}
\begin{Authlist}
\item \Idef{org1}A.I. Alikhanyan National Science Laboratory (Yerevan Physics Institute) Foundation, Yerevan, Armenia
\item \Idef{org2}Bogolyubov Institute for Theoretical Physics, National Academy of Sciences of Ukraine, Kiev, Ukraine
\item \Idef{org3}Bose Institute, Department of Physics  and Centre for Astroparticle Physics and Space Science (CAPSS), Kolkata, India
\item \Idef{org4}Budker Institute for Nuclear Physics, Novosibirsk, Russia
\item \Idef{org5}California Polytechnic State University, San Luis Obispo, California, United States
\item \Idef{org6}Central China Normal University, Wuhan, China
\item \Idef{org7}Centre de Calcul de l'IN2P3, Villeurbanne, Lyon, France
\item \Idef{org8}Centro de Aplicaciones Tecnol\'{o}gicas y Desarrollo Nuclear (CEADEN), Havana, Cuba
\item \Idef{org9}Centro de Investigaci\'{o}n y de Estudios Avanzados (CINVESTAV), Mexico City and M\'{e}rida, Mexico
\item \Idef{org10}Centro Fermi - Museo Storico della Fisica e Centro Studi e Ricerche ``Enrico Fermi', Rome, Italy
\item \Idef{org11}Chicago State University, Chicago, Illinois, United States
\item \Idef{org12}China Institute of Atomic Energy, Beijing, China
\item \Idef{org13}Comenius University Bratislava, Faculty of Mathematics, Physics and Informatics, Bratislava, Slovakia
\item \Idef{org14}COMSATS University Islamabad, Islamabad, Pakistan
\item \Idef{org15}Creighton University, Omaha, Nebraska, United States
\item \Idef{org16}Department of Physics, Aligarh Muslim University, Aligarh, India
\item \Idef{org17}Department of Physics, Pusan National University, Pusan, Republic of Korea
\item \Idef{org18}Department of Physics, Sejong University, Seoul, Republic of Korea
\item \Idef{org19}Department of Physics, University of California, Berkeley, California, United States
\item \Idef{org20}Department of Physics, University of Oslo, Oslo, Norway
\item \Idef{org21}Department of Physics and Technology, University of Bergen, Bergen, Norway
\item \Idef{org22}Dipartimento di Fisica dell'Universit\`{a} 'La Sapienza' and Sezione INFN, Rome, Italy
\item \Idef{org23}Dipartimento di Fisica dell'Universit\`{a} and Sezione INFN, Cagliari, Italy
\item \Idef{org24}Dipartimento di Fisica dell'Universit\`{a} and Sezione INFN, Trieste, Italy
\item \Idef{org25}Dipartimento di Fisica dell'Universit\`{a} and Sezione INFN, Turin, Italy
\item \Idef{org26}Dipartimento di Fisica e Astronomia dell'Universit\`{a} and Sezione INFN, Bologna, Italy
\item \Idef{org27}Dipartimento di Fisica e Astronomia dell'Universit\`{a} and Sezione INFN, Catania, Italy
\item \Idef{org28}Dipartimento di Fisica e Astronomia dell'Universit\`{a} and Sezione INFN, Padova, Italy
\item \Idef{org29}Dipartimento di Fisica `E.R.~Caianiello' dell'Universit\`{a} and Gruppo Collegato INFN, Salerno, Italy
\item \Idef{org30}Dipartimento DISAT del Politecnico and Sezione INFN, Turin, Italy
\item \Idef{org31}Dipartimento di Scienze e Innovazione Tecnologica dell'Universit\`{a} del Piemonte Orientale and INFN Sezione di Torino, Alessandria, Italy
\item \Idef{org32}Dipartimento Interateneo di Fisica `M.~Merlin' and Sezione INFN, Bari, Italy
\item \Idef{org33}European Organization for Nuclear Research (CERN), Geneva, Switzerland
\item \Idef{org34}Faculty of Electrical Engineering, Mechanical Engineering and Naval Architecture, University of Split, Split, Croatia
\item \Idef{org35}Faculty of Engineering and Science, Western Norway University of Applied Sciences, Bergen, Norway
\item \Idef{org36}Faculty of Nuclear Sciences and Physical Engineering, Czech Technical University in Prague, Prague, Czech Republic
\item \Idef{org37}Faculty of Science, P.J.~\v{S}af\'{a}rik University, Ko\v{s}ice, Slovakia
\item \Idef{org38}Frankfurt Institute for Advanced Studies, Johann Wolfgang Goethe-Universit\"{a}t Frankfurt, Frankfurt, Germany
\item \Idef{org39}Fudan University, Shanghai, China
\item \Idef{org40}Gangneung-Wonju National University, Gangneung, Republic of Korea
\item \Idef{org41}Gauhati University, Department of Physics, Guwahati, India
\item \Idef{org42}Helmholtz-Institut f\"{u}r Strahlen- und Kernphysik, Rheinische Friedrich-Wilhelms-Universit\"{a}t Bonn, Bonn, Germany
\item \Idef{org43}Helsinki Institute of Physics (HIP), Helsinki, Finland
\item \Idef{org44}High Energy Physics Group,  Universidad Aut\'{o}noma de Puebla, Puebla, Mexico
\item \Idef{org45}Hiroshima University, Hiroshima, Japan
\item \Idef{org46}Hochschule Worms, Zentrum  f\"{u}r Technologietransfer und Telekommunikation (ZTT), Worms, Germany
\item \Idef{org47}Horia Hulubei National Institute of Physics and Nuclear Engineering, Bucharest, Romania
\item \Idef{org48}Indian Institute of Technology Bombay (IIT), Mumbai, India
\item \Idef{org49}Indian Institute of Technology Indore, Indore, India
\item \Idef{org50}Indonesian Institute of Sciences, Jakarta, Indonesia
\item \Idef{org51}INFN, Laboratori Nazionali di Frascati, Frascati, Italy
\item \Idef{org52}INFN, Sezione di Bari, Bari, Italy
\item \Idef{org53}INFN, Sezione di Bologna, Bologna, Italy
\item \Idef{org54}INFN, Sezione di Cagliari, Cagliari, Italy
\item \Idef{org55}INFN, Sezione di Catania, Catania, Italy
\item \Idef{org56}INFN, Sezione di Padova, Padova, Italy
\item \Idef{org57}INFN, Sezione di Roma, Rome, Italy
\item \Idef{org58}INFN, Sezione di Torino, Turin, Italy
\item \Idef{org59}INFN, Sezione di Trieste, Trieste, Italy
\item \Idef{org60}Inha University, Incheon, Republic of Korea
\item \Idef{org61}Institut de Physique Nucl\'{e}aire d'Orsay (IPNO), Institut National de Physique Nucl\'{e}aire et de Physique des Particules (IN2P3/CNRS), Universit\'{e} de Paris-Sud, Universit\'{e} Paris-Saclay, Orsay, France
\item \Idef{org62}Institute for Nuclear Research, Academy of Sciences, Moscow, Russia
\item \Idef{org63}Institute for Subatomic Physics, Utrecht University/Nikhef, Utrecht, Netherlands
\item \Idef{org64}Institute of Experimental Physics, Slovak Academy of Sciences, Ko\v{s}ice, Slovakia
\item \Idef{org65}Institute of Physics, Homi Bhabha National Institute, Bhubaneswar, India
\item \Idef{org66}Institute of Physics of the Czech Academy of Sciences, Prague, Czech Republic
\item \Idef{org67}Institute of Space Science (ISS), Bucharest, Romania
\item \Idef{org68}Institut f\"{u}r Kernphysik, Johann Wolfgang Goethe-Universit\"{a}t Frankfurt, Frankfurt, Germany
\item \Idef{org69}Instituto de Ciencias Nucleares, Universidad Nacional Aut\'{o}noma de M\'{e}xico, Mexico City, Mexico
\item \Idef{org70}Instituto de F\'{i}sica, Universidade Federal do Rio Grande do Sul (UFRGS), Porto Alegre, Brazil
\item \Idef{org71}Instituto de F\'{\i}sica, Universidad Nacional Aut\'{o}noma de M\'{e}xico, Mexico City, Mexico
\item \Idef{org72}iThemba LABS, National Research Foundation, Somerset West, South Africa
\item \Idef{org73}Jeonbuk National University, Jeonju, Republic of Korea
\item \Idef{org74}Johann-Wolfgang-Goethe Universit\"{a}t Frankfurt Institut f\"{u}r Informatik, Fachbereich Informatik und Mathematik, Frankfurt, Germany
\item \Idef{org75}Joint Institute for Nuclear Research (JINR), Dubna, Russia
\item \Idef{org76}Korea Institute of Science and Technology Information, Daejeon, Republic of Korea
\item \Idef{org77}KTO Karatay University, Konya, Turkey
\item \Idef{org78}Laboratoire de Physique Subatomique et de Cosmologie, Universit\'{e} Grenoble-Alpes, CNRS-IN2P3, Grenoble, France
\item \Idef{org79}Lawrence Berkeley National Laboratory, Berkeley, California, United States
\item \Idef{org80}Lund University Department of Physics, Division of Particle Physics, Lund, Sweden
\item \Idef{org81}Nagasaki Institute of Applied Science, Nagasaki, Japan
\item \Idef{org82}Nara Women{'}s University (NWU), Nara, Japan
\item \Idef{org83}National and Kapodistrian University of Athens, School of Science, Department of Physics , Athens, Greece
\item \Idef{org84}National Centre for Nuclear Research, Warsaw, Poland
\item \Idef{org85}National Institute of Science Education and Research, Homi Bhabha National Institute, Jatni, India
\item \Idef{org86}National Nuclear Research Center, Baku, Azerbaijan
\item \Idef{org87}National Research Centre Kurchatov Institute, Moscow, Russia
\item \Idef{org88}Niels Bohr Institute, University of Copenhagen, Copenhagen, Denmark
\item \Idef{org89}Nikhef, National institute for subatomic physics, Amsterdam, Netherlands
\item \Idef{org90}NRC Kurchatov Institute IHEP, Protvino, Russia
\item \Idef{org91}NRC «Kurchatov Institute»  - ITEP, Moscow, Russia
\item \Idef{org92}NRNU Moscow Engineering Physics Institute, Moscow, Russia
\item \Idef{org93}Nuclear Physics Group, STFC Daresbury Laboratory, Daresbury, United Kingdom
\item \Idef{org94}Nuclear Physics Institute of the Czech Academy of Sciences, \v{R}e\v{z} u Prahy, Czech Republic
\item \Idef{org95}Oak Ridge National Laboratory, Oak Ridge, Tennessee, United States
\item \Idef{org96}Ohio State University, Columbus, Ohio, United States
\item \Idef{org97}Petersburg Nuclear Physics Institute, Gatchina, Russia
\item \Idef{org98}Physics department, Faculty of science, University of Zagreb, Zagreb, Croatia
\item \Idef{org99}Physics Department, Panjab University, Chandigarh, India
\item \Idef{org100}Physics Department, University of Jammu, Jammu, India
\item \Idef{org101}Physics Department, University of Rajasthan, Jaipur, India
\item \Idef{org102}Physikalisches Institut, Eberhard-Karls-Universit\"{a}t T\"{u}bingen, T\"{u}bingen, Germany
\item \Idef{org103}Physikalisches Institut, Ruprecht-Karls-Universit\"{a}t Heidelberg, Heidelberg, Germany
\item \Idef{org104}Physik Department, Technische Universit\"{a}t M\"{u}nchen, Munich, Germany
\item \Idef{org105}Politecnico di Bari, Bari, Italy
\item \Idef{org106}Research Division and ExtreMe Matter Institute EMMI, GSI Helmholtzzentrum f\"ur Schwerionenforschung GmbH, Darmstadt, Germany
\item \Idef{org107}Rudjer Bo\v{s}kovi\'{c} Institute, Zagreb, Croatia
\item \Idef{org108}Russian Federal Nuclear Center (VNIIEF), Sarov, Russia
\item \Idef{org109}Saha Institute of Nuclear Physics, Homi Bhabha National Institute, Kolkata, India
\item \Idef{org110}School of Physics and Astronomy, University of Birmingham, Birmingham, United Kingdom
\item \Idef{org111}Secci\'{o}n F\'{\i}sica, Departamento de Ciencias, Pontificia Universidad Cat\'{o}lica del Per\'{u}, Lima, Peru
\item \Idef{org112}St. Petersburg State University, St. Petersburg, Russia
\item \Idef{org113}Stefan Meyer Institut f\"{u}r Subatomare Physik (SMI), Vienna, Austria
\item \Idef{org114}SUBATECH, IMT Atlantique, Universit\'{e} de Nantes, CNRS-IN2P3, Nantes, France
\item \Idef{org115}Suranaree University of Technology, Nakhon Ratchasima, Thailand
\item \Idef{org116}Technical University of Ko\v{s}ice, Ko\v{s}ice, Slovakia
\item \Idef{org117}Technische Universit\"{a}t M\"{u}nchen, Excellence Cluster 'Universe', Munich, Germany
\item \Idef{org118}The Henryk Niewodniczanski Institute of Nuclear Physics, Polish Academy of Sciences, Cracow, Poland
\item \Idef{org119}The University of Texas at Austin, Austin, Texas, United States
\item \Idef{org120}Universidad Aut\'{o}noma de Sinaloa, Culiac\'{a}n, Mexico
\item \Idef{org121}Universidade de S\~{a}o Paulo (USP), S\~{a}o Paulo, Brazil
\item \Idef{org122}Universidade Estadual de Campinas (UNICAMP), Campinas, Brazil
\item \Idef{org123}Universidade Federal do ABC, Santo Andre, Brazil
\item \Idef{org124}University of Cape Town, Cape Town, South Africa
\item \Idef{org125}University of Houston, Houston, Texas, United States
\item \Idef{org126}University of Jyv\"{a}skyl\"{a}, Jyv\"{a}skyl\"{a}, Finland
\item \Idef{org127}University of Liverpool, Liverpool, United Kingdom
\item \Idef{org128}University of Science and Technology of China, Hefei, China
\item \Idef{org129}University of South-Eastern Norway, Tonsberg, Norway
\item \Idef{org130}University of Tennessee, Knoxville, Tennessee, United States
\item \Idef{org131}University of the Witwatersrand, Johannesburg, South Africa
\item \Idef{org132}University of Tokyo, Tokyo, Japan
\item \Idef{org133}University of Tsukuba, Tsukuba, Japan
\item \Idef{org134}Universit\'{e} Clermont Auvergne, CNRS/IN2P3, LPC, Clermont-Ferrand, France
\item \Idef{org135}Universit\'{e} de Lyon, Universit\'{e} Lyon 1, CNRS/IN2P3, IPN-Lyon, Villeurbanne, Lyon, France
\item \Idef{org136}Universit\'{e} de Strasbourg, CNRS, IPHC UMR 7178, F-67000 Strasbourg, France, Strasbourg, France
\item \Idef{org137}Universit\'{e} Paris-Saclay Centre d'Etudes de Saclay (CEA), IRFU, D\'{e}partment de Physique Nucl\'{e}aire (DPhN), Saclay, France
\item \Idef{org138}Universit\`{a} degli Studi di Foggia, Foggia, Italy
\item \Idef{org139}Universit\`{a} degli Studi di Pavia, Pavia, Italy
\item \Idef{org140}Universit\`{a} di Brescia, Brescia, Italy
\item \Idef{org141}Variable Energy Cyclotron Centre, Homi Bhabha National Institute, Kolkata, India
\item \Idef{org142}Warsaw University of Technology, Warsaw, Poland
\item \Idef{org143}Wayne State University, Detroit, Michigan, United States
\item \Idef{org144}Westf\"{a}lische Wilhelms-Universit\"{a}t M\"{u}nster, Institut f\"{u}r Kernphysik, M\"{u}nster, Germany
\item \Idef{org145}Wigner Research Centre for Physics, Budapest, Hungary
\item \Idef{org146}Yale University, New Haven, Connecticut, United States
\item \Idef{org147}Yonsei University, Seoul, Republic of Korea
\end{Authlist}
\endgroup
\end{document}